\documentclass[12pt]{article}
\usepackage{epsfig,amssymb,amsmath,psfrag}

\def \be {\begin{equation}}
\def \ee {\end{equation}}
\def \ba  {\begin{eqnarray}}
\def \ea  {\end{eqnarray}}
\def \baa {\begin{eqnarray*}}
\def \eaa {\end{eqnarray*}}
\def \bb  {\begin {thebibliography} }
\def \eb  {\end{thebibliography}}
\def \lab #1 {\label{#1}}

\newcommand\re[1]{(\ref{#1})}

\def \qqqquad {\qquad\qquad}
\def \matrix #1 {\left(\begin{array}{cc} #1 \end{array}\right)}
\def \Tr {\mathop{\rm Tr}\nolimits}
\def \tr {\mathop{\rm tr}\nolimits}

\newcommand\lr[1]{{\left({#1}\right)}}

\newcommand \vev [1] {\langle{#1}\rangle}
\newcommand \VEV [1] {\left\langle{#1}\right\rangle}

\newcommand{\ft}[2]{{\textstyle\frac{#1}{#2}}}
\def \vep {\epsilon}

\def\m{\mu}
\def\C{\Gamma}

\def\nn{\nonumber}

\newcommand{\n}{\nu}
\renewcommand{\l}{\lambda}
\newcommand{\pa}{\partial}
\newcommand{\cN}{{\cal N}}

\newcommand{\ep}{\epsilon}
\newcommand{\p}[1]{(\ref{#1})}

\begin{document}

\thispagestyle{empty}
\null\vskip-12pt \hfill  LAPTH-1224/07 \\
\null\vskip-12pt \hfill LPT--Orsay--07--133
\vskip2.2truecm
\begin{center}
\vskip 0.2truecm {\Large\bf
{\Large  Conformal Ward identities for Wilson loops and\\[2mm]  a test of
  the duality with gluon amplitudes}
}\\
\vskip 1truecm
{\bf J.M. Drummond$^{*}$, J. Henn$^{*}$, G.P. Korchemsky$^{**}$ and E. Sokatchev$^{*}$ \\
}

\vskip 0.4truecm
$^{*}$ {\it Laboratoire d'Annecy-le-Vieux de Physique Th\'{e}orique
LAPTH\footnote{UMR 5108 associ\'{e}e \`{a}
 l'Universit\'{e} de Savoie},\\
B.P. 110,  F-74941 Annecy-le-Vieux, France\\
\vskip .2truecm $^{**}$ {\it
Laboratoire de Physique Th\'eorique%
\footnote{Unit\'e Mixte de Recherche du CNRS (UMR 8627)},
Universit\'e de Paris XI, \\
F-91405 Orsay Cedex, France
                       }
  } \\
\end{center}

\vskip 1truecm 
\centerline{\bf Abstract} \normalsize 

\medskip

\noindent
Planar gluon amplitudes in $\cN=4$ SYM are remarkably similar to expectation
values of Wilson loops made of light-like segments. We argue that the latter can
be determined by making use of the conformal symmetry of the gauge theory, broken
by cusp anomalies. We derive the corresponding anomalous conformal Ward
identities valid to all loops and show that they uniquely fix the form of the
finite part of a Wilson loop with $n$ cusps (up to an additive constant) for
$n=4$ and 5 and reduce the freedom in it to a function of conformal invariants
for $n\geq6$. We also present an explicit two-loop calculation for $n=5$. The
result confirms the form predicted by the Ward identities and matches the
finite part of the two-loop five-gluon planar MHV amplitude,  up to a constant. This constitutes
another non-trivial test of the Wilson loop/gluon amplitude duality.

\newpage

\setcounter{page}{1}\setcounter{footnote}{0}

\section{Introduction}

The recent surge of interest in gluon scattering amplitudes in $\mathcal{N}=4$
SYM theory has been motivated by two findings. The four-gluon color-ordered
planar amplitudes reveal an intriguing iterative structure at weak coupling --
the Bern-Dixon-Smirnov (BDS) conjecture \cite{bds05}, which predicts the finite
part of the amplitude in terms of the cusp anomalous dimensions~\cite{P80,KR87}
and {two other functions of the coupling~\cite{Anastasiou:2003kj}.} Quite
remarkably, the same structure also emerged at strong coupling within the
Alday-Maldacena proposal \cite{am07} for a string description of the gluon
scattering amplitude in the AdS/CFT correspondence~\cite{AdS}. Their string
construction makes contact with a Wilson loop on a specific contour with cusps
and like-like edges determined by the gluon momenta. Such light-like Wilson loops
have already been studied in \cite{KK92,Kr02}.

 Inspired by these recent developments, in \cite{DKS07} three of us conjectured
that a new type of duality between gluon amplitudes and Wilson loops may also exist at weak coupling
in $\mathcal{N}=4$ SYM theory with the gauge group $SU(N)$.
More precisely, the conjectured duality relates the following two objects.
The first one, ${\mathcal{M}_4}$, is the color-ordered partial planar four-gluon amplitude divided by its tree-level expression.
The second one is the planar limit of the expectation value of the Wilson
loop

\begin{equation}\label{1steq}
W(C_4)=\frac1{N}\vev{0|{\rm Tr}\, {\rm P} \exp\lr{i\oint_{C_4} dx\cdot A(x)}|0}\,,
 \end{equation}
where the integration
contour  $C_4$  consists of four light-like segments
$[x_i,x_{i+1}]$ in a dual Minkowski space-time with coordinates $x_i^\mu$ related to the on-shell gluon momenta,
\footnote{Note that when considering gluon amplitudes,
we fix the canonical dimension of length to be $-1$.
When considering Wilson loops, we fix the canonical dimension of length to be $+1$.
This explains why coordinates in one picture can be related to momenta in the other picture.
This clearly shows that dual coordinates are not identified with the position space of the gluons.
 }
\begin{equation}\label{xpidentificationintro}
{p_i^\mu := x_i^\mu-x_{i+1}^\mu  } \,.
\end{equation}
Both objects have, respectively, infrared and ultraviolet divergences, $\ln {\mathcal{M}_4} = \ln Z_4 + \ln F_4 + O(\epsilon_{\rm IR})$
and $\ln W(C_4) = \ln Z_4^{\rm (WL)} + \ln F_4^{\rm (WL)} + O(\epsilon_{\rm UV})$, where $Z$ contains up to second-order
poles in the infrared and ultraviolet dimensional regulators $\epsilon_{\rm IR}$ and $\epsilon_{\rm UV}$, respectively, and $F$ denotes
the finite part as the regulator is taken to zero.
While the relation between the divergent parts of light-like Wilson loops and gluon amplitudes is well known \cite{KR87},
the non-trivial content of the duality conjectured in \cite{DKS07} is the matching of the
{\it finite} parts of the two objects,
\begin{equation}\label{M=W}
\ln F_4 = \ln F^{\rm (WL)}_4 + {\rm const}   \,.
\end{equation}
In (\ref{M=W}) the kinematic variables describing the scattering amplitude are identified with the
distances appearing in the Wilson loop according to (\ref{xpidentificationintro}).
This is the weak coupling version of the strong coupling duality demonstrated by Alday and Maldacena.
At weak coupling, we have shown by an explicit one-loop
\cite{DKS07} and then two-loop calculation \cite{Drummond:2007cf} that the relation \re{M=W}
reproduces the known expression for the four-gluon amplitude \cite{Bern:1997nh,Anastasiou:2003kj}.

The BDS conjecture \cite{bds05} not only applies to four-gluon amplitudes but
also to $n-$gluon maximal helicity violating (MHV) planar amplitudes. So far the
conjecture has been confirmed for $n=4$ up to three loops in \cite{bds05}, while
for $n>4$ it has been tested only for $n=5$ and at two loops in \cite{5point}.
The Alday-Maldacena proposal~\cite{am07,Alday:2007he} also works for an arbitrary
number of gluons (for generalization to other processes see
Refs.~\cite{Komargodski:2007er,McGreevy:2007kt,Ito:2007zy}), although the
practical evaluation of the corresponding solution of the classical string
equations is difficult even for $n=5$ {(for a recent discussion see
Refs.~\cite{Kruczenski:2007cy,Mironov:2007qq,Ryang:2007bc,Astefanesei:2007bk,Mironov:2007xp,Yang:2007cm,Itoyama:2007ue})}.
However, Alday and Maldacena \cite{Alday:2007he} proposed a way to do this for
very large $n$, which gives a disagreement with the BDS conjecture at strong
coupling. It is straightforward to generalize the weak-coupling duality relation
\re{M=W} to MHV amplitudes with an arbitrary number $n\ge 5$ of external legs on
the one hand, and the expectation value of a Wilson loop evaluated along a
polygonal contour consisting of $n$ light-like segments, on the other hand. This
duality has been tested for arbitrary $n$ at one loop in \cite{BHT07}.

In \cite{Drummond:2007cf} we argued that we can profit from the (broken)
conformal symmetry of the light-like Wilson loop to make all-loop predictions
about the form of the amplitude.\footnote{It should be noted that conformal
arguments were efficiently used by Alday and Maldacena to obtain their
strong-coupling minimal surface solution in \cite{am07}.} Due to the presence of a cusp anomaly in
the Wilson loop, conformal invariance manifests itself in the form of anomalous
Ward identities. We proposed a very simple anomalous conformal Ward identity for
the finite part of the amplitude and conjectured it to be valid to all orders in
the coupling. This identity uniquely fixes the form of the finite part (up to an
additive constant) of the Wilson loop dual to the four- and five-gluon
amplitudes, and gives partial restrictions on the functional dependence on the
kinematic variables for $n\geq6$. Quite remarkably, the BDS ansatz \cite{bds05}
for the $n-$gluon MHV amplitudes satisfies the conformal Ward identity for
arbitrary $n$ \cite{Drummond:2007cf}.

If the duality relation \re{M=W} holds for any $n$ and to all loops, the
conformal symmetry properties of the light-like Wilson loops impose strong
constraints on the finite part of the gluon amplitudes in $\mathcal{N}=4$ SYM.
This may provide the natural explanation of the BDS conjecture for $n=4$ and
$n=5$, but cannot validate it for $n\geq6$. One might also try to find a direct
conformal symmetry argument on the MHV amplitude side. The encouraging fact is
that the $n=4$ gluon amplitudes have been shown to possess a peculiar `hidden' or
`dual' conformal symmetry \cite{Drummond:2006rz} (different from the conformal
symmetry of the underlying $\cN=4$ theory) up to four loops \cite{Bern:2006ew}
(and even at five loops \cite{Bern:2007ct}). It is natural to conjecture that the
controlled breaking of this symmetry in the on-shell divergent amplitude will
lead to the same type of Ward identities.

In this paper we present a derivation
of the all-order anomalous conformal Ward identities conjectured in
\cite{Drummond:2007cf}. In addition, we perform an explicit two-loop calculation
of the Wilson loop made out of five light-like segments (pentagon). We show that
the two-loop expression for the pentagon Wilson loop indeed satisfies the
conformal Ward identities. Most importantly, it coincides with the known
expression~\cite{5point} for the two-loop correction to the five-gluon planar
scattering amplitude, thus providing additional support for the gluon
amplitude/Wilson loop correspondence \re{M=W}.

The presentation is organized as follows. In Section \ref{section-generalities}
we review some properties of light-like Wilson loops which will be important in
deriving the conformal Ward identities in section \ref{section-CWI}. Then, in
section \ref{section-pentagon}, we present a two-loop calculation of
the pentagon Wilson loop as an explicit check of our Ward identities.

\section{General features of light-like Wilson
loops}\label{section-generalities}

\subsection{Definitions}

The central object of our consideration is the light-like Wilson loop defined in
$\mathcal{N}=4$ SYM theory with $SU(N)$ gauge group as
\begin{equation}\label{W}
    W\lr{C_n} = \frac1{N}\vev{0|\,{\rm Tr}\, \textrm{P} \exp\lr{i\oint_{C_n} dx^\mu A_\mu(x)}
    |0}\,,
\end{equation}
where $ A_\mu(x)=A_\mu^a(x) t^a$ is a gauge field, $t^a$ are the $SU(N)$
generators in the fundamental representation normalized as $\tr (t^a t^b)=\ft12
\delta^{ab}$ and $\textrm{P}$ indicates the ordering of the $SU(N)$ indices along
the integration contour $C_n$. This contour is made out of $n$ light-like
segments  $C_n=\bigcup_{i=1}^n \ell_i$ joining the cusp points $x_i^\mu$ (with
$i=1,2,\ldots,n$)
\begin{equation}\label{4'}
 \ell_i=\{x^\mu(\tau_i)= \tau_i x^\mu_i +
(1-\tau_i)x_{i+1}^\mu|\, \tau_i \in[0,1] \}\,,
\end{equation}
such that the vectors
\begin{equation}\label{0002}
   { x_{i,i+1}^\mu \equiv x_{i}^\mu - x_{i+1}^\mu :=p_i^\mu}\,, \qquad p^2_i=0
\end{equation}
are identified with the external on-shell momenta of the $n$-gluon scattering
amplitude. Thinking about $W\lr{C_n}$ as a function of the cusp points we observe
that it has the following symmetry
\begin{align}\label{cyclic}
W(x_1,x_2,\ldots,x_n) = W(x_n,x_1,\ldots,x_{n-1}) = W(x_n,x_{n-1},\ldots,x_1)\,,
\end{align}
where the first relation is a consequence the cyclic property of the trace on the
right-hand side of \re{W}, while the second relation follows from reality
condition $\lr{W(C_n)}^* = W\lr{C_n}$.

\subsection{Cusp singularities}

A distinctive feature of the contour $C_n$ is the presence of $n$ cusps located
at the points $x_i^\mu$. This causes specific ultraviolet divergences (UV) to
appear in the Wilson loop~\cite{P80}. The fact that the cusp edges are light-like
makes them even more severe~\cite{KK92}. In order to get some insight into the
structure of these divergences we start by giving a one-loop example. Later on in
this subsection we give arguments which lead to the all-order form of the
divergences.

\subsubsection{One-loop example}

\begin{figure}
\psfrag{xi1}[cc][cc]{$x_{i}$} \psfrag{xi3}[cc][cc]{$x_{i-1}$}
\psfrag{xi2}[cc][cc]{$x_{i+1}$} \psfrag{x}[cc][cc]{$x$}
\psfrag{a}[cc][cc]{(a)} \psfrag{b}[cc][cc]{(b)}
\psfrag{c}[cc][cc]{(c)}
\centerline{\includegraphics[height=50mm,keepaspectratio]{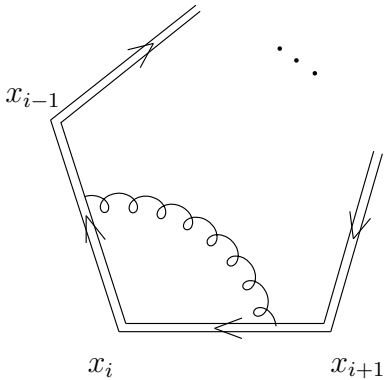}}
\caption[]{\small The Feynman diagram contributing to the one-loop divergence at the cusp point $x_i$. The double
line depicts the integration contour $C_n$, the wiggly line the
gluon propagator.}
\label{Fig:5gon}
\end{figure}

Let us first discuss the origin of the cusp singularities in the polygonal Wilson
loop $W(C_n)$ to the lowest order   in the coupling constant. According to
definition \re{W}, it is given by a double contour integral,
\begin{equation}\label{double-sum}
 W(C_n) = 1 + \frac12(ig)^2 C_F
 \oint_{C_n}  dx^\mu \oint_{C_n} dy^\nu\, G_{\mu\nu}(x-y)  +
 O(g^4)  \,.
\end{equation}
Here $C_F= (N^2-1)/(2N)$ is the quadratic Casimir of $SU(N)$ in the
fundamental representation and $G_{\mu\nu}(x-y)$ is the gluon propagator in the
coordinate representation
\begin{equation}\label{a1}
   \vev{A_\m^a(x)\ A_\nu^b(y)}= g^2 \delta^{ab} G_{\mu\nu}(x-y) \,.
\end{equation}
To regularize the ultraviolet divergences of the integrals entering
\re{double-sum}, we shall employ dimensional regularization
\footnote{In fact, we use the dimensional reduction scheme in order
to preserve supersymmetry. This will become essential when we generalize
to higher loops in the next subsection.},
$D=4-2\epsilon$ (with $\epsilon>0$). Also,  making use of the gauge
invariance of the Wilson loop \re{W} and for the sake of simplicity, we
perform the calculation in the Feynman gauge, where the gluon
propagator is given by \footnote{\label{foot1} As in
\cite{Drummond:2007cf}, we redefine the conventional dimensional
regularization scale as $ \mu^2\pi{\rm e}^\gamma\mapsto\mu^2$ to
avoid dealing with factors involving $\pi$ and the Euler constant
$\gamma$.}
\begin{equation}\label{propagator}
G_{\mu\nu}(x)  = g_{\mu\nu} G(x)\,,\qquad G(x)= - \frac{\Gamma(1-\vep)}{4\pi^{2}}
(- x^2+i0)^{-1+\vep}\lr{\mu^2 {\rm e}^{-\gamma}}^{\vep}\,.
\end{equation}
The divergences in \re{double-sum} originate from the integration of the position
of the gluon in the vicinity of a light-like cusp in the diagram shown in
Fig.~\ref{Fig:5gon}.
The calculation of this diagram is straightforward and the details can be found
e.g. in \cite{KK92}. Adding together the contributions of all cusps we obtain the
following one-loop expression for the divergent part of $W(C_n)$:
\begin{equation}\label{1-loop}
\ln W(C_n) = \frac{g^2}{4\pi^2}C_F \left\{ -\frac{1}{2\vep^2}  \sum_{i=1}^n
\lr{{-x_{i-1,i+1}^2}\,{\mu^2}}^\vep + O(\vep^0)\right\}+O(g^4)\,,
\end{equation}
where the periodicity condition
\begin{equation}
x_{ij}^2=x_{i+n,j}^2=x_{i,j+n}^2 \equiv (x_i-x_j)^2
\end{equation}
is tacitly implied.

\subsubsection{All-loop structure}\label{section-Wnall-loop}

Generalizing \re{1-loop} to higher loops we find that the cusp singularities
appear in $W(C_n)$ to the $l$-th loop-order as poles $W(C_n)\sim (a
\mu^{2\epsilon})^l/\epsilon^m$ with $m\le 2l$. Furthermore, $W(C_n)$ can be split
into a divergent (`renormalization') factor $Z_n$ and a finite (`renormalized')
factor $F_n$ as
\begin{equation}\label{W=ZF}
W(C_n) = Z_n F_n\ .
\end{equation}
{}From the studies of renormalization properties of light-like Wilson loops it is
known~\cite{KK92} that cusp singularities exponentiate to all loops and, as a
consequence, the factor $Z_n$ has the special form \footnote{ Formula (\ref{5})
follows from the evolution equation (8) of the first reference in \cite{KK92}.}
\begin{equation}\label{5}
    \ln Z_n = -\frac{1}{4}  \sum_{l\ge 1} a^l\sum_{i=1}^n\lr{-x_{i-1,i+1}^2\mu^2}^{l\ep} \lr{\frac{\Gamma_{\rm
cusp}^{(l)}}{(l\ep)^2}+ \frac{\Gamma^{(l)}}{l\ep}} \,, \qquad a=\frac{g^2
N}{8\pi^2}\ .
\end{equation}
Here $\Gamma_{\rm cusp}^{(l)}$ and $\Gamma^{(l)}$ are the expansion coefficients
of the cusp anomalous dimension and the so-called collinear anomalous dimension,
respectively, defined in the {adjoint} representation of $SU(N)$:
\begin{align}\label{cusp-2loop}
& \Gamma_{\rm cusp}(a)=\sum_{l\ge 1} a^l\, \Gamma_{\rm cusp}^{(l)} = 2a -
\frac{\pi^2}3 a^2 + O(a^3)\ ,
\\ \notag
& \Gamma(a) =\sum_{l\ge 1} a^l\, \Gamma^{(l)} = - 7\zeta_3 a^2 + O(a^3)\,.
\end{align}

We have already seen at one-loop order that the divergences in $W(C_n)$ are due
to the presence of the cusps on the integration contour $C_n$. They occur when
the gluon propagator in Fig.~\ref{Fig:5gon} slides along the contour towards a
given cusp point $x_i$. The divergences appear when the propagator becomes
singular. This happens either when a gluon propagates at short distances in the
vicinity of the cusp point (short distance divergences), or when a gluon
propagates along the light-like segment adjacent to the cusp point (collinear
divergences). In these two regimes we encounter, respectively, a double and
single pole.

Going to higher orders in the coupling expansion of $W(C_n)$, we immediately
realize that the structure of divergences coming from each individual diagram
becomes much more complicated. Still, the divergences originate from the same
part of the `phase space' as at one-loop: from short distances in the vicinity of
cusps and from propagation along light-like edges of the contour. The main
difficulty in analyzing these divergences is due to the fact that in diagrams
with several gluons attached to different segments of $C_n$ various regimes could
be realized simultaneously, thus enhancing the strength of poles in $\epsilon$.
This implies, in particular, that individual diagrams could generate higher order
poles to $W(C_n)$.

The simplest way to understand the form of $\ln Z_n$ in (\ref{5}) is
to analyze the divergences of the Feynman diagrams not in the Feynman gauge
but in the axial gauge defined as
\begin{equation}
n \cdot A(x) = 0
\end{equation}
with $n^\mu$ being an arbitrary vector, $n^2 \neq 0$. The reason for this is that
the same Feynman diagrams become less singular in the axial gauge and, most
importantly, the potentially divergent graphs have a much simpler
topology~\cite{DDT}.\footnote{The contribution of individual Feynman diagrams is
gauge dependent and it is only their total sum that is gauge invariant.} The
axial gauge gluon propagator in the momentum representation is given by the
following expression
\begin{equation}
\widetilde G_{\mu\nu}(k) = -i\frac{d_{\mu\nu}(k)}{k^2+i0}\,,\qquad
d_{\mu\nu}^{(A)}(k)=
 {g_{\mu\nu} - \frac{k_\mu n_\nu + k_\nu n_\mu}{(k n)}+ k_\mu k_\nu
\frac{n^2}{(k  n)^2}}\,.
\end{equation}
The polarization tensor satisfies
the relation
\begin{equation}g^{\mu\nu} d^{(A)}_{\mu\nu}(k) = (D-2) +
\frac{k^2n^2}{(k n)^2}
\end{equation}
(to be compared with the
corresponding relation in the Feynman gauge, which is $g^{\mu\nu}
d_{\mu\nu}^{(F)}(k)=D$) from which we deduce that for $k^2=0$, it
describes only the
physical polarizations of the on-shell gluon.%
\footnote{That is the reason why the axial gauge is called physical.} Let us
consider a graph in which a gluon is attached to the $i$-th segment. In
configuration space, the corresponding effective vertex is described by the
contour integral $\int d\tau_i\, p_i^\mu\, A_\mu(x_i-p_i\tau_i)$. In the momentum
representation, the same vertex reads $\int d\tau_i\, p_i^\mu\, \tilde A_\mu(k)\,
{\rm e}^{ik(x_i-p_i\tau_i) }$ where the field $\tilde A_\mu(k)$ describes all
possible polarizations ($2$ longitudinal and $D-2$ transverse) of the gluon with
momentum $k^\mu$. Let us examine the collinear regime, when the gluon propagates
along the light-like direction $p^\mu_i$. The fact that the gluon momentum is
collinear, $k^\mu \sim p^\mu_i$, implies that it propagates close to the
light-cone and, therefore, has a small virtuality $k^2$. Furthermore, since
$p_i^\mu \tilde A_\mu(k) \sim k^\mu \tilde A_\mu(k)$ we conclude that the
contribution of the transverse polarization of the gluon is suppressed as
compared to the longitudinal ones. In other words, the most singular contribution
in the collinear limit comes from the longitudinal components of the gauge field
$\tilde A_\mu(k)$. The properties of the latter depend on the gauge, however. To
see this, let us examine the form of the emission vertex in the Feynman and in
the axial gauge. In the underlying Feynman integral, the gauge field $\tilde
A_\mu(k)$ will be replaced by the propagator $\tilde D_{\mu\nu}(k)$, with $\nu$
being the polarization index at the vertex to which the gluon is attached. In
this way, we find that
\begin{equation}
p_i^\mu\, \tilde G_{\mu\nu}(k) \sim k^\mu \tilde G_{\mu\nu}(k) = -i\frac{k^\mu
d_{\mu\nu}(k)}{k^2+i0} = -\frac{i}{k^2}\times \bigg\{\begin{array}{ll}
  k^\nu\,, & \text{Feynman gauge} \\
  k^2\left[\frac{k^\nu n^2}{(kn)^2}-\frac{n^\nu}{(kn)}\right], & \text{axial gauge} \\
\end{array}
\end{equation}
Since $k^2\to 0$ in the collinear limit, we conclude that the vertex is
suppressed in the axial gauge, as compared to the Feynman gauge. This is in
perfect agreement with our physical intuition -- the propagation of longitudinal
polarizations of a gluon with momentum $k^\mu$ is suppressed for $k^2\to 0$. This
property is rather general and it holds not only for gluons attached to the
integration contour, but also for the `genuine' interaction vertices of the
$\mathcal{N}=4$ SYM Lagrangian~\cite{DDT}. It should not be surprising now that
the collinear divergences in the light-like Wilson loop come from graphs of very
special topology that we shall explain in a moment.

Another piece of information that will be extensively used in our analysis comes
from the so-called non-Abelian exponentiation property of Wilson
loops~\cite{Gatheral}. It follows from the combinatorial properties of the
path-ordered exponential and it is not sensitive to the particular form of the
Lagrangian of the underlying gauge theory. For an arbitrary integration contour
$C$ it can be formulated as follows:
\begin{equation}\label{exponentiation}
\vev{W(C)} = 1+ \sum_{k=1}^\infty
\lr{\frac{g^2}{4\pi^2}}^k W^{(k)} = \exp\left[{\sum_{k=1}^\infty
\lr{\frac{g^2}{4\pi^2}}^k c^{(k)} w^{(k)}}\right]\,.
\end{equation}
Here $W^{(k)}$ denote the perturbative corrections to the Wilson loop, while
$c^{(k)}w^{(k)}$ are given by the contribution to $W^{(k)}$ from `webs' $w^{(k)}$
with the `maximally non-Abelian' color factor $c^{(k)}$. To the first few orders,
$k=1,2,3$, the maximally non-Abelian color factor takes the form $c^{(k)}=C_F
N^{k-1}$, but starting from $k=4$ loops it is not expressible in terms of simple
Casimir operators. Naively, one can think of the `webs' $w^{(k)}$ as of Feynman
diagrams with maximally interconnected gluon lines. For the precise definition of
`webs' we refer the interested reader to \cite{Gatheral}.

\begin{figure}
\psfrag{xi1}[cc][cc]{$x_{i}$} \psfrag{xi3}[cc][cc]{$x_{i-1}$}
\psfrag{xi2}[cc][cc]{$x_{i+1}$} \psfrag{x}[cc][cc]{$x$}
\psfrag{a}[cc][cc]{(a)} \psfrag{b}[cc][cc]{(b)}
\psfrag{c}[cc][cc]{(c)}
\centerline{\includegraphics[height=50mm,keepaspectratio]{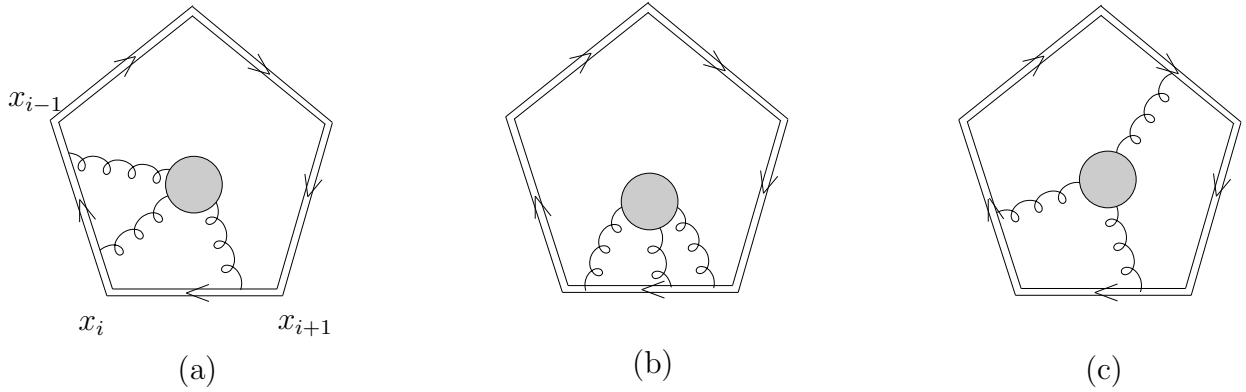}}
\caption[]{\small {Maximally non-abelian Feynman diagrams of different topologies
(`webs') contributing to $\ln W(C_n)$. In the axial gauge, the vertex-type
diagram (a) generates a simple pole; the self-energy type diagram (b) generates a
double pole in $\epsilon$; diagram (c) with gluons attached to three and more
segments is finite.} }
\label{Fig:webs}
\end{figure}

Let us now return to the analysis of the cusp divergences of the light-like
Wilson loops and take advantage of both the axial gauge and the exponentiation
\re{exponentiation}. A characteristic feature of the `webs' following from their
maximal non-Abelian nature is that the corresponding Feynman integrals have
`maximally complicated' momentum loop flow.
When applied to the light-like Wilson loop, this has the
following remarkable consequences in the axial gauge~\cite{KK92}:
\begin{itemize}
  \item the `webs' do not contain nested divergent subgraphs;
  \item each web produces a double pole in $\epsilon$ at most;
  \item the divergent contribution only comes from
`webs' localized at the cusp points as shown in Fig.~\ref{Fig:webs}(a)
    and (b).
\end{itemize}

These properties imply that the divergent part of $\ln W(C_n)$ is given by a sum
over the cusp points $x_i^\mu$ with each cusp producing a double and a single pole
contribution~\cite{KK92}. Moreover, the corresponding residues depend on $x_{i-1,i+1}^2$ --
the only kinematic invariant that one can build out of the three vectors
$x_{i-1}^\mu$, $x_i^\mu$ and $x_{i+1}^\mu$, satisfying
$x_{i-1,i}^2=x_{i,i+1}^2=0$. In this way, we arrive at the known relation \re{5}
for the divergent part of the light-like Wilson loop.

Our consideration relied on the analysis of Feynman diagrams at weak coupling. It
was recently shown in Refs.~\cite{Kr02,Buch07,AM07mf} that the structure of
divergences of $\ln W(C_n)$ remains the same even at strong coupling.

\section{Conformal symmetry of light-like Wilson loops }\label{section-CWI}

Were the Wilson loop $W\lr{C_n}$ well defined in $D=4$ dimensional Minkowski
space-time, then it would enjoy the (super)conformal invariance of the underlying
$\mathcal{N}=4$ SYM theory. More precisely, in the absence of conformal anomalies
we would conclude that
\begin{equation}\label{cin}
    W\lr{C'_n} = W\lr{C_n}\,,
\end{equation}
where $C'_n$ is the image of the contour $C_n$ upon an $SO(2,4)$ conformal
transformations.

Our contour $C_n$ is very special, however. Firstly, its shape is stable under
conformal transformations, that is the contour $C'_n$ is also made of $n$
light-like segments with new cusp points ${x_i'}^{\mu}$ obtained as the images of
the old ones $x_i^\mu$. This property is rather obvious for translations,
rotations and dilatations but we have to check it for special conformal
transformations. Performing a conformal inversion~\footnote{A special conformal
transformation (boost) is equivalent to an inversion followed by a translation
and then another inversion.}, ${x^\mu}' = {x^\mu}/{x^2}$, of all points belonging
to the segment \re{4'}, we obtain another segment of the same type,
${x'}^\mu(\tau_i')= \tau_i' {x'}_i^\mu + (1-\tau_i') {x'}_{i+1}^\mu$ with
$(x'_{i+1}-x'_i)^2=0$ and $\tau_i' = \tau_i/[\tau_i+
(1-\tau_i)(x'_{i})^2/(x'_{i+1})^2]$.

The second distinctive feature of the contour $C_n$ is the presence of cusps,
which, as we already pointed out in section \ref{section-generalities}, cause
specific ultraviolet divergences to appear in the Wilson loop (\ref{W}). For this
reason we employ dimensional regularization with $D=4-2\epsilon$ (with
$\epsilon>0$).

In the dimensionally regularized $\mathcal{N}=4$ SYM theory, the Wilson loop
$W\lr{C_n}\equiv \vev{W_n}$ is given by a functional integral
\begin{align}\label{W-path-integral}
\vev{W_n} =  \int {\cal D}A\, {\cal D}\lambda \, {\cal D}\phi\ {\rm e}^{i
S_{\epsilon}[A,\,\lambda,\,\phi] }\ {\rm Tr}\left[ \textrm{P} \exp\left(i \oint_{C_n}
dx\cdot A(x) \right)\right]\,,
\end{align}
where the integration goes over gauge fields, $A$, gaugino, $\lambda$, and scalars \footnote{We remind the reader that
we use the supersymmetry preserving DRED scheme, in which there are $6+2\epsilon$ scalars.},
$\phi$, with the action
\begin{equation}\label{3}
    S_{\epsilon}= \frac{1}{g^2\mu^{2\ep}}\int d^{D}x\ {\cal L}(x)\,, \qquad
    {\cal L} = \mbox{Tr} { \left[-\ft{1}{2}F^2_{\mu\nu}\right]} + \mbox{gaugino}+\text{scalars}+\text{gauge
    fixing}+\text{ghosts}.
\end{equation}
Here $\m$ is the regularization scale and we redefined all fields in such a way
that $g$ does not appear inside the Lagrangian ${\cal L}(x)$. This allows us to
keep the {\it canonical} dimension of all fields, in particular of the gauge
field $A^\mu(x)$, and hence to preserve the conformal invariance of the
path-ordered exponential entering the functional integral in
\re{W-path-integral}. However, due to the change of dimension of the measure
$\int d^D x$  in (\ref{3}) the action $S_\epsilon$ is not invariant under
dilatations and conformal boosts, which yields an {\it anomalous} contribution to
the Ward identities.

\subsection{Anomalous conformal Ward identities}

The conformal Ward identities for the light-like Wilson loop $W(C_n)$ can be
derived following the standard method
\cite{Sarkar:1974xh,Mueller:1993hg,Braun:2003rp}. To begin with we recall the
well-known expressions for the generators of the $SO(2,4)$ conformal
transformations -- rotations ($\mathbb{M}^{\m\n}$), dilatations ($\mathbb{D}$),
translations ($\mathbb{P}^\mu$) and special conformal boosts ($\mathbb{K}^\m$),
acting on fundamental (gauge, gaugino, scalars) fields $\phi_I(x)$ with conformal
weight $d_\phi$ and Lorentz indices $I$~\footnote{The generators $\mathbb{G}$
determine the infinitesimal transformations with parameters $\varepsilon$:
$\phi'(x) = \phi(x) + \varepsilon\cdot \mathbb{G} \phi(x)$.}
\begin{eqnarray}
  &\mathbb{M}^{\m\n}  \phi_I &= (x^\m \pa^\n - x^\n \pa^\m)  \phi_I + (m^{\m\n})_I{}^J \phi_J \,,  \nn\\
  &\mathbb{D} \,\phi_I &= x\cdot\pa\ \phi_I + d_\phi\, \phi_I  \,,  \nn  \\
  &\mathbb{P}^\m  \phi_I &= \pa^\m\ \phi_I \,, \nn\\
  &\mathbb{K}^\m \phi_I &= \left(2 x^\m x\cdot\pa -  x^2 \pa^\m \right) \phi_I + 2x^\m \, d_\phi\, \phi_I + 2x_\nu (m^{\m\n})_I{}^J \phi_J \ ,  \label{1'}
\end{eqnarray}
where $m^{\m\n}$ is the generator of spin rotations, e.g., $m^{\m\n}=0$ for a
scalar field and $(m^{\m\n})_\l{}^\rho = g^{\n\rho} \delta^\m_\l - g^{\m\rho}
\delta^\n_\l$ for a gauge field.

Let us start with the dilatations and perform a change of variables in the
functional integral \re{W-path-integral}, $\phi_I'(x) = \phi_I(x)
+\varepsilon\,\mathbb{D} \phi_I(x)$. This change of variables could be
compensated by a coordinate transformation ${x^{\mu}}' = (1-\varepsilon)x^\m$. We
recall that the path-ordered exponential is invariant under dilatations, whereas
the Lagrangian is covariant with canonical weight $d_{\cal L}=4$. However, the
measure $\int d^Dx$ with $D=4-2\ep$ does not match the weight of the Lagrangian,
which results in a non-vanishing variation of the action $S_\epsilon$
\begin{equation} \delta_\mathbb{D} S_\epsilon = -\frac{2\ep}{g^2\mu^{2\ep}}\int
d^{D}x\ {\cal L}(x)\,.
\end{equation}
This variation generates an operator insertion into the
expectation value, $\vev{\delta_\mathbb{D} S_\epsilon\ W_n}$, and yields an
anomalous term in the action of the dilatation generator on $\vev{W_n}$
\begin{equation}\label{6}
    \mathbb{D} \vev{W_n} = \sum_{i=1}^n \lr{x_i\cdot \pa_i} \ \vev{W_n} =
     {-\frac{2i\ep}{g^2\mu^{2\ep}}}\int d^{D}x\ \langle{\cal L}(x) W_n \rangle\ .
\end{equation}
In a similar manner,  the anomalous special conformal (or conformal boost) Ward
identity is derived by performing transformations generated by the operator
$\mathbb{K}^\nu$, Eq.~(\ref{1'}), on both sides of \re{W-path-integral}. In this
case the nonvanishing variation of the action $\delta_{\mathbb{K}^\mu}
S_\epsilon$ comes again from the mismatch of the conformal weights of the
Lagrangian and of
the measure $\int d^D x$.%
\footnote{Another source of non-invariance of the action $S_\epsilon$ is the
gauge-fixing term (and the associated ghost term of the non-Abelian theory) which
is not conformally invariant even in four dimensions. However, due to gauge
invariance of the Wilson loop, such anomalous terms do not appear on the
right-hand side of (\ref{20}).} This amounts to considering just the $d_\phi$
term in (\ref{1'}) with $d_\phi = d_{\cal L} - D = 2\ep$, hence
\begin{equation}\label{20}
     \mathbb{K}^\nu \vev{W_n} = \sum^n_{i=1} (2x_i^\nu x_i\cdot\pa_i - x_i^2 \pa_i^\nu) \vev{W_n}
     = - \frac{4i\ep}{g^2\mu^{2\ep}}\int d^{D}x\ x^\nu\ \langle {\cal L}(x) W_n \rangle \ .
\end{equation}
The relations \re{6} and \re{20} can be rewritten as
\begin{align}\label{21}
& \mathbb{D} \ln\vev{W_n} =
     -\frac{2i\ep}{g^2\mu^{2\ep}}\int d^{D}x\ \frac{\langle{\cal L}(x) W_n \rangle}{\vev{W_n}}\
     ,
\\ \notag
&    \mathbb{K}^\nu \ln\vev{W_n}   = - \frac{4i\ep}{g^2\mu^{2\ep}}\int d^{D}x\
x^\nu\
    \frac{\langle{\cal L}(x) W_n \rangle}{\vev{W_n}}\ .
\end{align}
To make use of these relations we have to evaluate the ratio $\langle{\cal L}(x)
W_n \rangle/\vev{W_n}$ obtained by inserting the Lagrangian into the Wilson loop
expectation value. Due to the presence of $\epsilon$ on the right-hand side of
\re{21}, it is sufficient to know its divergent part only.

\subsection{Dilatation Ward identity}

As we will now show, the dilatation Ward identity can be derived by dimensional
arguments and this provides a consistency condition for the right-hand side of
(\ref{21}). By definition \re{W-path-integral}, the dimensionally regularized
light-like Wilson loop $\vev{W_n}$ is a dimensionless scalar function of the cusp
points $x_i^\nu$ and, as a consequence, it satisfies the relation
\begin{equation}\label{8}
     \left(\sum_{i=1}^n (x_i\cdot \pa_i) -\mu \frac{\pa}{\pa \mu}   \right) \ln\vev{W_n}= 0\ .
\end{equation}
In addition, its perturbative expansion is expressed in powers of the coupling
$g^2\mu^{2\ep}$ and, therefore,
\begin{equation}\label{7}
  \mu \frac{\pa}{\pa \mu} \vev{W_n} =   2\ep g^2 \frac{\pa}{\pa g^2} \vev{W_n}
  = -\frac{2i \ep}{g^2\mu^{2\ep}}\int d^{D}x\ \langle{\cal L}(x) W_n \rangle\ ,
\end{equation}
where the last relation follows from \re{W-path-integral}.

{The Wilson loop $\vev{W_n}$ can be split into the product of divergent and
finite parts, Eq.~\re{W=ZF}}.  Notice that the definition of the divergent part
is ambiguous as one can always add to $\ln Z_n$ a term finite for $\epsilon\to
0$. Our definition \re{5} is similar to the conventional $\rm MS$ scheme with the
only difference that we choose the expansion parameter to be $a\mu^{2\epsilon}$
instead of $a$. The reason for this is that $Z_n$ satisfies, in our scheme, the
same relation $\mu {\pa}_\mu Z_n = 2\ep g^2 {\pa}_{g^2} Z_n $ as $\vev{W_n}$
\p{7}. Together with \p{W=ZF} and \re{8}, this implies that the finite part of
the Wilson loop does not depend on the renormalization point, i.e.
\begin{equation}\label{Fn-indep-eps}
    \mu{\pa_\mu} F_n = O(\epsilon)\,.
\end{equation}
Writing $\ln\vev{W_n} = \ln Z_n + \ln F_n$, and using the explicit
form of $Z_n$ in (\ref{5}), the relation (\ref{7}) leads to the following dilatation Ward identity%
\footnote{In what follows, we shall systematically neglect corrections to $F_n$
vanishing as $\epsilon\to 0$.}
\begin{equation}\label{F-dilatations}
 \sum_{i=1}^n (x_i\cdot \pa_{x_i}) \, F_n= 0\,.
\end{equation}
Adding to this the obvious requirement of Poincar\'{e}
invariance, we conclude that the finite part ${F}_n$ of the light-like Wilson
loop can only depend on the dimensionless ratios $x^2_{ij}/x^2_{kl}$. In
particular, for $n=4$ there is only one such independent ratio, i.e. ${F}_4 =
{F}_4\left( {x^2_{13}}/{x^2_{24}}\right)$.

Making use of \re{W=ZF}, \re{5} and \re{F-dilatations} we find that the all-loop
dilatation Ward identity for $\vev{W_n}$ takes the form
\begin{equation}\label{D-all-loop}
\mathbb{D} \ln \vev{W_n}= \sum_{i=1}^n
(x_i\cdot \pa_i) \, \ln\vev{W_n} = -\frac{1}{2} \sum_{l\ge 1}
a^l\sum_{i=1}^n\lr{-x_{i-1,i+1}^2\mu^2}^{l\ep} \lr{\frac{\Gamma_{\rm
cusp}^{(l)}}{l\ep}+ {\Gamma^{(l)}}} \ .
\end{equation}
This relation provides a constraint on the form of the Lagrangian insertion on
the right-hand side of (\ref{21}).

\subsection{One-loop calculation of the anomaly}

The derivation of the dilatation Ward identities \re{F-dilatations} relied on the
known structure of cusp singularities (\ref{5}) of the light-like Wilson loop and
did not require a detailed knowledge of the properties of Lagrangian insertion
$\langle{\cal L}(x) W_n \rangle/\vev{W_n}$. This is not the case anymore for the
special conformal Ward identity.

\begin{figure}
\psfrag{xi1}[cc][cc]{$x_{i}$} \psfrag{xi3}[cc][cc]{$x_{i-1}$}
\psfrag{xi2}[cc][cc]{$x_{i+1}$} \psfrag{x}[cc][cc]{$x$}
\psfrag{a}[cc][cc]{(a)} \psfrag{b}[cc][cc]{(b)}
\psfrag{c}[cc][cc]{(c)}
\centerline{\includegraphics[height=50mm,keepaspectratio]{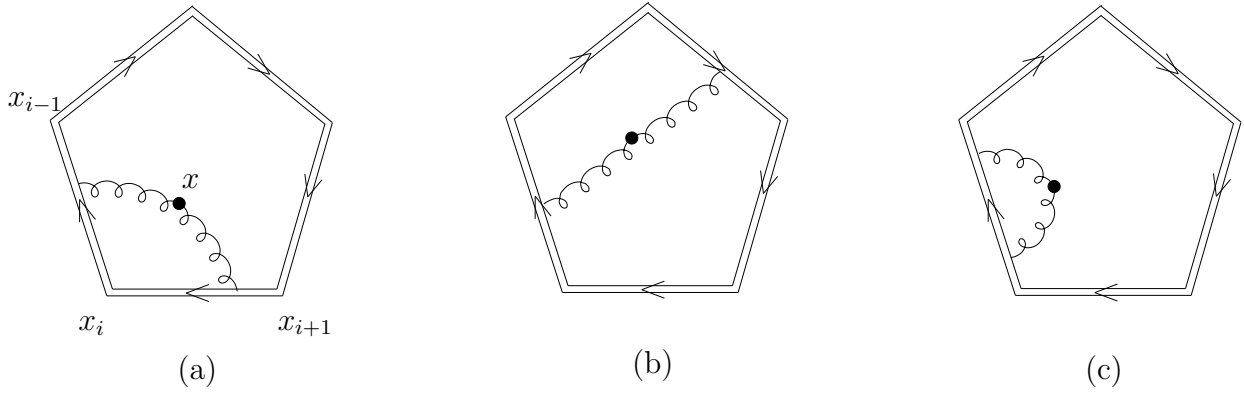}}
\caption[]{\small The Feynman diagrams contributing to $\langle{\cal L}(x) W_n
\rangle$ to the lowest order in the coupling. The double line depicts the
integration contour $C_n$, the wiggly line the gluon propagator and the blob the
insertion point.}
\label{Fig:insertion}
\end{figure}

To start with, let us perform an explicit one-loop computation of $\langle{\cal
L}(x) W_n \rangle/\vev{W_n}$. To the lowest order in the coupling, we substitute
$\vev{W_n}=1 +O(g^2)$ and retain inside ${\cal L}(x)$ and $W_n$ only terms
quadratic in gauge field. The result is
\begin{equation}\label{insertion}
\frac{\langle{\cal L}(x) W_n \rangle}{\vev{W_n}} = {-\frac{1}{4N}\VEV{\mbox{Tr}
\left[(\partial_\mu A_\nu(x) - \partial_\nu A_\mu(x))^2\right]\Tr\left[ \big(i{\mbox{$\oint_{C_n}$} dy \cdot
A(y)}\big)^2\right] }+ O(g^6)}\,.
\end{equation}
The Wick
contractions between gauge fields coming from the Lagrangian and the path-ordered
exponential yield a product of two gluon propagators \re{a1}, each connecting the
point $x$ with an arbitrary point $y$ on the integration contour $C_n$. Gauge
invariance allows us to choose, e.g., the Feynman gauge in which the gluon
propagator is given by \re{propagator}. To the lowest order in the coupling
constant, the right-hand side of \re{insertion} receives non-vanishing
contributions only from Feynman diagrams of three different topologies shown in
Figs.~\ref{Fig:insertion}(a)  -- (c).

In the Feynman gauge, only the vertex-like diagram shown in
Fig.~\ref{Fig:insertion}(a)
develops poles in $\epsilon$. Performing the calculation, we find after some algebra%
\footnote{It is advantageous to perform the calculation by taking a Fourier
transform with respect to $x$ and later take the inverse transform of the final
result.}
\begin{align}\label{fin222}
 {\frac{2i}{g^2\mu^{2\ep}}\frac{\langle{\cal L}(x) W_n \rangle}{\vev{W_n}}}  =  a \sum_{i=1}^n
\lr{-x_{i-1,i+1}^2\mu^2}^\ep\bigg\{\ep^{-2}\delta^{(D)}(x-x_i)+\ep^{-1} \Upsilon^{(1)}(x;
x_{i-1},x_i,x_{i+1})+ O(\epsilon^0)\bigg\},
\end{align}
where $a=g^2N/(8\pi^2)$ and we introduced the notation
\begin{equation}
\Upsilon^{(1)}(x; x_{i-1},x_i,x_{i+1}) =\int_0^1 \frac{ds}{s}
\bigg[{\delta^{(D)}(x-x_i- s x_{i-1,i})}+{\delta^{(D)}(x-x_i+ s
x_{i,i+1})-2\delta^{(D)}(x-x_i)} \bigg].
\end{equation}We see that the leading
double-pole singularities are localized at the cusp points $x=x_i$. The
subleading single poles are still localized on the contour, but they are
`smeared' along the light-like edges adjacent to the cusp.

Substitution of \re{fin222} into \re{21} yields
\begin{align}\label{CWI-1loop}
& \mathbb{D} \ln\vev{W_n} = -a \sum_{i=1}^n \lr{-x_{i-1,i+1}^2\mu^2}^\ep \ep^{-1}
+O(a^2)
     ,
\\ \notag
&    \mathbb{K}^\nu \ln\vev{W_n}   = -2a \sum_{i=1}^n
x_i^\nu\lr{-x_{i-1,i+1}^2\mu^2}^\ep \ep^{-1}  +O(a^2) \ .
\end{align}
Notice that $\sum_{i=1}^n \Upsilon^{(1)}(x; x_{i-1},x_i,x_{i+1})$ does not
contribute to the right-hand sides of these relations by virtue of
\begin{align}\label{Y-identity}
& \int d^D x\, \Upsilon^{(1)}(x; x_{i-1},x_i,x_{i+1}) = 0\,,
\\ \notag
& \int d^D x\,x^\nu \Upsilon^{(1)}(x; x_{i-1},x_i,x_{i+1}) =
(x_{i-1}+x_{i+1}-2x_i)^\nu\,.
\end{align}
As was already mentioned, the right-hand sides of the Ward identities
\re{CWI-1loop} are different from zero due to the fact that the light-like Wilson
loop has cusp singularities.

We verify with the help of \re{cusp-2loop} that to the lowest order in the
coupling, the first relation in \re{CWI-1loop} is in agreement with
\re{D-all-loop}.

\subsection{Structure of the anomaly to all loops}

To extend the analysis of the special conformal Ward identity to all loops we
examine the all-loop structure of the divergences of $\langle{\cal L}(x) W_n
\rangle/\vev{W_n}$. They arise in a way very similar to those of the Wilson loop
itself, which were discussed in section \ref{section-Wnall-loop}.

It is convenient to couple ${\cal L}(x)$ to an auxiliary `source' $J(x)$ and
rewrite the insertion of the Lagrangian into $\vev{W_n}$ as a functional
derivative
\begin{equation}
\langle{\cal L}(x) W_n \rangle/\vev{W_n} =
-i \frac{\delta }{\delta J(x)}\ln \vev{W_n}_J\bigg|_{J=0},
\end{equation}
where the subscript $J$ indicates that the expectation value is taken in the
$\mathcal{N}=4$ SYM theory with the additional term $\int d^D x J(x){\cal L}(x)$
added to the action. This generates new interaction vertices inside the Feynman
diagrams for $\vev{W_n}_J$ but does not affect the non-Abelian
exponentiation property \re{exponentiation}.%
\footnote{We remark that non-Abelian exponentiation is a `kinematic' property of the Wilson loop $W_n$ in the sense that it follows from the very definition of
$W_n$ as a path-ordered exponential, combined with the combinatorial properties of the $SU(N_c)$ generators \cite{Gatheral}. As a consequence, it is insensitive to the form of the action.}
The only difference compared with \re{exponentiation} is
that the webs $w^{(k)}$ now depend on $J(x)$ through a new interaction vertex.
Making use of non-Abelian exponentiation, we obtain
\begin{equation}\langle{\cal L}(x) W_n \rangle/\vev{W_n} = {\sum_{k=1}^\infty
\lr{\frac{g^2}{4\pi^2}}^k c^{(k)} \left[-i\frac{\delta w^{(k)}}{\delta
J(x)}\right] \bigg|_{J=0} }\,.
\label{jderiv}
\end{equation}
Similarly to $\ln \vev{W_n}$, the webs produce at most double poles in $\epsilon$
localized at a given cusp. This allows us to write a general expression for the
divergent part of the Lagrangian insertion:
\begin{eqnarray}\label{L-insertion}
&& \frac{2i\ep}{g^2\mu^{2\ep}} \ \frac{\langle{\cal L}(x) W_n \rangle}{\vev{W_n}} \label{25} \\
   &=& \ \sum_{l\ge 1} a^l \sum_{i=1}^n \lr{-x_{i-1,i+1}^2\mu^2}^{l\ep}
   \left\{ \frac12\lr{\frac{\Gamma_{\rm cusp}^{(l)}}{ l\ep
}+  \Gamma^{(l)}}\delta^{(D)}(x-x_i)  + \Upsilon^{(l)}(x; x_{i-1},x_i,x_{i+1})
\right\} + O(\ep)\ ,  \nn
\end{eqnarray}
which generalizes \re{fin222} to all loops. The following comments are in order.

In Eq.~\re{L-insertion}, the term proportional to $\delta^{(D)}(x-x_i)$ comes
from the double pole contribution to the web $w^{(k)}$ which is indeed located at
the short distances in the vicinity of the cusp $x_i^\mu$. The residue of the
simple pole in the right-hand side of \re{L-insertion} is given by the cusp
anomalous dimension. This can be shown by substituting \re{L-insertion} into the
dilatation Ward identity \p{6} and comparing with (\ref{D-all-loop}).

The contact nature of the leading singularity in \p{L-insertion} can also be
understood in the following way. The correlator on the left-hand side can be
viewed as a conformal $(n+1)$-point function with the Lagrangian at one point and
the rest corresponding to the cusps. In the $\cN=4$ SYM theory the Lagrangian
belongs to the protected stress-tensor multiplet, therefore it has a fixed
conformal dimension four. The Wilson loop itself, if it were not divergent, would
be conformally invariant. This means that the $n$ cusp points can be regarded as
having vanishing conformal weights. Of course, the presence of divergences might
make the conformal properties anomalous. However, the  conformal behaviour of the
leading singularity in \p{L-insertion} cannot be corrected by an
anomaly.\footnote{Such an anomalous contribution should come from a $1/\ep^3$
pole in the correlator with two insertions of the Lagrangian, but repeated use of
the argument above shows that the order of the poles does not increase with the
number of insertions.} Then we can argue that the only function of space-time
points, which has conformal weight four at one point and zero at all other
points, is the linear combination of delta functions appearing in
\p{L-insertion}.\footnote{The mismatch of the conformal weight four of the
Lagrangian and $D=4-2\ep$ of the delta functions does not affect the leading
singularity in \p{L-insertion}.}

As was already emphasized, the residue of the simple pole of $\delta
w^{(k)}/\delta J(x)$  at the cusp point $x_i$ depends on its position and at most
on its two nearest neighbors $x_{i-1}$ and $x_{i+1}$, as well as on the insertion
point~\cite{KK92}. It gives rise to a function $\Upsilon^{(l)}(x; x_{i-1},x_i,x_{i+1})$,
which is the same for all cusp points due to the cyclic symmetry of the Wilson
loop \re{cyclic}.

Notice that in \p{25} we have chosen to separate the terms with the collinear
anomalous dimension $\Gamma^{(l)}$ from the rest of the finite terms. This choice
has implications for the function $\Upsilon^{(l)}(x; x_{i-1},x_i,x_{i+1})$.
Substituting the known factor $Z_n$ \p{5} and the particular form of \p{25} into
the dilatation Ward identity \p{6}, we derive
\begin{equation}\label{26}
    \sum_{i=1}^n \int d^Dx\ \Upsilon^{(l)}(x; x_{i-1},x_i,x_{i+1}) = 0\ .
\end{equation}
We can argue that in fact each term in this sum vanishes. Indeed, each term in
the sum is a Poincar\'{e} invariant dimensionless function of three points
$x_{i-1}$, $x_i$ and $x_{i+1}$. Given the light-like separation of the
neighboring points, the only available invariant is $x_{i-1,i+1}^2$. Further,
$\Upsilon^{(l)}(x; x_{i-1},x_i,x_{i+1})$ cannot depend on the regularization
scale because $\m$ always comes in the combination $a\m^{2\ep}$ and thus
contributes to the $O(\ep)$ terms in \p{25}. The dimensionless Poincar\'{e}
invariant $\int d^Dx\ \Upsilon^{(l)}(x; x_{i-1},x_i,x_{i+1})$ depends on a single
scale and, therefore, it must be a constant after which \p{26} implies
\begin{equation}\label{26'}
    \int d^Dx\ \Upsilon^{(l)}(x; x_{i-1},x_i,x_{i+1}) = 0\ .
\end{equation}
For $l=1$ this relation is in agreement with the one-loop result \re{Y-identity}.

\subsection{Special conformal Ward identity}

We are now ready to investigate the special conformal Ward identity \p{21}.
Inserting \p{25} into the right-hand side and integrating over $x$ we obtain
\begin{eqnarray}
  \mathbb{K}^\nu \ln W_n &=& \sum^n_{i=1} (2x_i^\nu x_i\cdot\pa_i - x_i^2 \pa_i^\nu) \ln W_n
  \label{28}\\
  &=& -\sum_{l\ge 1} a^l \lr{\frac{\Gamma_{\rm cusp}^{(l)}}{ l\ep
}+  {\Gamma^{(l)}} }\sum_{i=1}^n\ \lr{-x_{i-1,i+1}^2\mu^2}^{l\ep} \ x^\nu_i  -2
\sum_{i=1}^n \Upsilon^\nu(x_{i-1},x_i,x_{i+1}) + O(\ep) \ ,   \nn
\end{eqnarray}
where
\begin{equation}\label{29}
    \Upsilon^\nu(x_{i-1},x_i,x_{i+1})  = \sum_{l\ge 1}a^l \int d^D x\, x^\nu \Upsilon^{(l)}(x; x_{i-1},x_i,x_{i+1})\ .
\end{equation}
Next, we substitute $\ln W_n =\ln Z_n + \ln F_n$ into \p{28}, replace $\ln Z_n$
by its explicit form \p{5} and expand the right-hand side in powers of $\ep$ to
rewrite \p{28} as follows:
\begin{equation}\label{32}
    \mathbb{K}^\nu \ln {F}_n  =  \frac{1}{2} \Gamma_{\rm cusp}(a) \sum_{i=1}^n
    \ln \frac{x_{i,i+2}^2}{x_{i-1,i+1}^2} x^\nu_{i,i+1} - 2 \sum_{i=1}^n
    \Upsilon^\nu(x_{i-1},x_i,x_{i+1}) + O(\ep)\ .
\end{equation}
Note that the quantities $\Upsilon^\nu(x_{i-1},x_i,x_{i+1})$ are translation
invariant. Indeed, a translation under the integral in \p{29} only affects the
factor $x^\nu$ (the functions $\Upsilon^{(l)}(x; x_{i-1},x_i,x_{i+1})$ are
translation invariant), but the result vanishes as a consequence of \p{26'}.
Furthermore, $\Upsilon^\nu(x_{i-1},x_i,x_{i+1})$ only depends on two neighboring
light-like vectors $x^\m_{i-1,i}$ and $x^\m_{i,i+1}$, from which we can form only
one non-vanishing Poincar\'{e} invariant, $x^2_{i-1,i+1}$. We have already argued
that $\Upsilon^{(l)}$ are independent of $\mu$, and so must be $\Upsilon^\nu$.
Taking into account the scaling dimension one of $\Upsilon^\nu$, we conclude that
\begin{equation}\label{34}
    \Upsilon^\nu(x_{i-1},x_i,x_{i+1})  = \alpha x^\nu_{i-1,i} + \beta x^\nu_{i,i+1}\ ,
\end{equation}
where $\alpha,\ \beta $ only depend on the coupling. The symmetry of the Wilson
loop $W_n$ under mirror exchange of the cusp points, Eq.~\re{cyclic}, translates
into symmetry of \re{34} under exchange of the neighbors $x_{i-1}$ and $x_{i+1}$
of the cusp point $x_i$, which reduces (\ref{34}) to
\begin{equation}\label{35}
    \Upsilon^\nu(x_{i-1},x_i,x_{i+1})   = \alpha \, (x^\nu_{i-1} + x^\nu_{i+1} - 2 x^\nu_i)\
    ,
\end{equation}
with $\alpha=a + O(a^2)$ according to \re{Y-identity}.
 Substituting this relation into \re{32} we find
\begin{equation}\label{36}
    \sum_{i=1}^n \Upsilon^\nu(x_{i-1},x_i,x_{i+1}) = 0\ .
\end{equation}

This concludes the derivation of the special conformal Ward identity. In the
limit $\ep\to0$ it takes the form (cf. \cite{Drummond:2007cf})
\begin{equation}\label{37}
    \sum^n_{i=1} (2x_i^\nu x_i\cdot\pa_i - x_i^2 \pa_i^\nu) \ln {F}_n  =  \frac{1}{2} \Gamma_{\rm cusp}(a) \sum_{i=1}^n  \ln \frac{x_{i,i+2}^2}{x_{i-1,i+1}^2} x^\nu_{i,i+1}\ .
\end{equation}

\subsection{Solution and implications for $W_n$}\label{crrrr}

Let us now examine the consequences of the conformal Ward identity \p{37} for the
finite part of the Wilson loop $W_n$. We find that the cases of $n=4$ and $n=5$
are special because here the Ward identity \p{37} has a unique solution up to an
additive constant. The solutions are, respectively,
\begin{align}
\ln {F}_4 &= \frac{1}{4}\Gamma_{\rm cusp}(a)
\ln^2\Bigl(\frac{x_{13}^2}{x_{24}^2}\Bigr) + \text{
  const }\,, \nn\\
\ln {F}_5 &= - \frac{1}{8}\Gamma_{\rm cusp}(a) \sum_{i=1}^5 \ln
  \Bigl(\frac{x_{i,i+2}^2}{x_{i,i+3}^2}\Bigr) \ln
  \Bigl(\frac{x_{i+1,i+3}^2}{x_{i+2,i+4}^2}\Bigr) + \text{ const }\, , \label{remarkable}
\end{align}
as can be easily verified by making use of the identity  {$\mathbb{K}^{\mu}
x_{ij}^2 = 2 (x_i^\mu + x_j^\mu) x_{ij}^2 $}. We find that, upon identification
of the kinematic invariants
\begin{equation}\label{xx} x_{k,k+r}^2 := (p_k+\ldots + p_{k+r-1})^2\,, \end{equation}the
relations (\ref{remarkable}) are exactly the functional forms of the ansatz of
\cite{bds05} for the finite parts of the four- and five-point MHV amplitudes (or
rather the ratio of the amplitude to the corresponding tree amplitude).

The reason why the functional form of ${F}_4$ and ${F}_5$ is fixed up to an
additive constant is that there are no conformal invariants one can build from
four or five points $x_i$ with light-like separations $x_{i,i+1}^2=0$. Such
invariants take the form of cross-ratios\footnote{{For $n$ generic points
$x^\mu_i$, $i=1\ldots n$ in a $D$-dimensional space-time the number of invariants
is $(n-1)(n-2)/2 - 1$ if $n\leq D+1$ and $nD - (D+1)(D+2)/2$ if $n > D+1$. The
additional conditions of light-like separations, $x^2_{i,i+1}=0$, remove $n$ of
them.}}
\begin{equation}\label{crr}
    \frac{x^2_{ij}x^2_{kl}}{x^2_{ik}x^2_{jl}}\ .
\end{equation}
It is obvious that with four or five points they cannot be constructed. This
becomes possible starting from six points, where there are three such
cross-ratios, e.g., \begin{equation}u_1 = \frac{x_{13}^2 x_{46}^2}{x_{14}^2
x_{36}^2}, \qquad u_2 = \frac{x_{24}^2 x_{15}^2}{x_{25}^2 x_{14}^2}, \qquad u_3 =
\frac{x_{35}^2 x_{26}^2}{x_{36}^2 x_{25}^2}\ . \end{equation} Hence the general
solution of the Ward identity at six cusp points and higher will contain an
arbitrary function of the conformal cross-ratios.

\section{Light-like pentagon Wilson loop}\label{section-pentagon}

In this section we perform an explicit two-loop calculation of the light-like
pentagon Wilson loop $W(C_5)$ in $\mathcal{N}=4$ SYM theory. It goes along the
same lines as the analysis of the rectangular ($n=4$) {and hexagon ($n=6$)
Wilson loops performed in
Refs.~\cite{Drummond:2007cf,KK92,Drummond:2008aq},} where the interested reader can find the
details of the technique employed.

\subsection{One loop result}

The one-loop calculation of $\ln W(C_5)$ was done in \cite{BHT07}.
The relevant Feynman diagrams are shown in Fig.~\ref{all-diags}(a)
and Fig.~\ref{all-diags}(b).

Let us split $\ln W(C_5)$ into divergent and finite parts
\begin{equation}
\label{W=div+fin} \ln W(C_5) =\frac{g^2}{4\pi^2}C_F \left[D^{(1)} +
F^{(1)}\right] + O(g^4)\,,
\end{equation}
where by definition
\begin{align}\label{D,F}
D^{(1)} &= -\frac{1}{2\vep^2} \sum_{i=1}^5 \lr{{-x_{i,i+2}^2}\,{\mu^2}}^\vep\,,
\\\notag
F^{(1)} &= - \frac{1}{4} \sum_{i=1}^5 \ln
 \Bigl(\frac{x_{i,i+2}^2}{x_{i,i+3}^2}\Bigr) \ln
 \Bigl(\frac{x_{i+1,i+3}^2}{x_{i+2,i+4}^2}\Bigr) + \frac{5\pi^2}{24} + O(\vep)\,.
\end{align}
{The pentagon Wilson loop \re{W=div+fin} fulfills a duality relation generalizing the four-point one \re{M=W} (see \cite{BHT07}).} To
verify this,
we apply \re{0002} to
identify the coordinates $x_{i,i+1}^\mu$ with the on-shell gluon momenta
$p_i^\mu$. This leads to
\begin{equation}
x_{i,i+2}^2 := (p_i+p_{i+1})^2 \equiv s_{i,i+1}\,,\qquad
x_{i,i+3}^2 = x_{i,i-2}^2:= (p_{i-2}+p_{i-1})^2 \equiv s_{i-1,i-2}\,,
\end{equation}
where $s_{jk} = (p_j+p_k)^2$ are the Mandelstam invariants corresponding to the
five-gluon amplitude. A specific feature of the $n=5$ on-shell gluon amplitude as
compared with $n\ge 6$ is that it depends only on two-particle invariants
$s_{i,i+1}$. Similarly, for the light-like Wilson loop one finds that $W(C_5)$
only depends on the distances $x_{i,i+2}^2$ between next-to-neighboring vertices
on the contour $C_5$.
Then we observe that, firstly, upon identification of the
dimensional regularization parameters
\begin{equation}\label{1-loop-match}
\epsilon = - \epsilon_{\rm IR}\,,\qquad x_{i,i+2}^2\,\mu^2 := s_{i,i+1}/\mu_{\rm
IR}^2 \,,\qquad x_{i,i+2}^2/x_{k,k+2}^2 := s_{i,i+1}/s_{k,k+1}\,,
\end{equation}
the UV divergences of the light-like Wilson loop match the IR divergent part of
the five-gluon scattering amplitude and, secondly, the finite corrections to
these two objects indeed coincide at one loop,  up to an additive constant.

In the next section, we extend the analysis beyond one loop and demonstrate that
the planar $n=5$ gluon amplitude/pentagon Wilson loop duality also holds to two
loops.

\subsection{Two-loop calculation}

As was explained in detail in \cite{Drummond:2007cf}, the two-loop calculation of
$W(C_5)$ can be significantly simplified by making use of the non-Abelian
exponentiation property of Wilson loops~\cite{Gatheral}. In application to
$W(C_5)$, it can be formulated as follows:
\begin{equation}\label{W-decomposition}
\ln W(C_5) = \frac{g^2}{4\pi^2}C_F\,
w^{(1)}  +  \lr{\frac{g^2}{4\pi^2}}^2 C_F N\, w^{(2)}  + O(g^6)\,,
\end{equation}
where $w^{(1)}$ and $w^{(2)}$ are functions of the distances
$x_{i,i+2}$, independent of the Casimirs of the gauge group $SU(N)$. Matching
\re{W-decomposition} into \re{W=div+fin} we find that
\begin{equation}
w^{(1)} = D^{(1)} + F^{(1)}\,,
\end{equation}
with $D^{(1)}$ and $F^{(1)}$ defined in
\re{D,F}. The relation \re{exponentiation} implies that the coefficient in front
of $g^4C_F^2/(4\pi^2)^2$ in the two-loop expression for the pentagon Wilson loop
$W(C_5)$ is given by $\frac12\lr{w^{(1)}}^2$ and, therefore, it is uniquely
determined by the one-loop correction to $W(C_5)$. Thus, in order to determine
the function $w^{(2)}$ it sufficient to calculate the contribution to $W(C_5)$
only from two-loop diagrams containing `maximally non-Abelian' color factor $C_F
N$. This property allows us to reduce significantly the number of relevant
two-loop diagrams. In addition, as yet another advantage of using the Feynman
gauge, we observe \cite{Drummond:2007cf} that some of the `maximally non-Abelian'
diagrams like those where  both ends of a gluon are attached to the same
light-like segment vanish by virtue of $x_{j,j+1}^2 =0$. The corresponding
Feynman diagrams have the same topology as for the rectangular Wilson loop
$W(C_4)$ and they can be easily identified by applying the selection rules
formulated in Ref.~\cite{Drummond:2007cf}.

\begin{figure}
\psfrag{x1}[cc][cc]{$x_3$} \psfrag{x2}[cc][cc]{$x_2$} \psfrag{x3}[cc][cc]{$x_1$}
\psfrag{x4}[cc][cc]{$x_5$} \psfrag{x5}[cc][cc]{$x_4$}
\psfrag{a}[cc][cc]{(a)} \psfrag{b}[cc][cc]{(b)} \psfrag{c}[cc][cc]{(c)}
\psfrag{d}[cc][cc]{(d)} \psfrag{e}[cc][cc]{(e)} \psfrag{f}[cc][cc]{(f)}
\psfrag{g}[cc][cc]{(g)} \psfrag{h}[cc][cc]{(h)} \psfrag{i}[cc][cc]{(i)}
\psfrag{j}[cc][cc]{(j)} \psfrag{k}[cc][cc]{(k)} \psfrag{l}[cc][cc]{(l)}
\psfrag{m}[cc][cc]{(m)} \psfrag{n}[cc][cc]{(n)} \psfrag{o}[cc][cc]{(o)}
%
\centerline{\includegraphics[height=200mm,keepaspectratio]{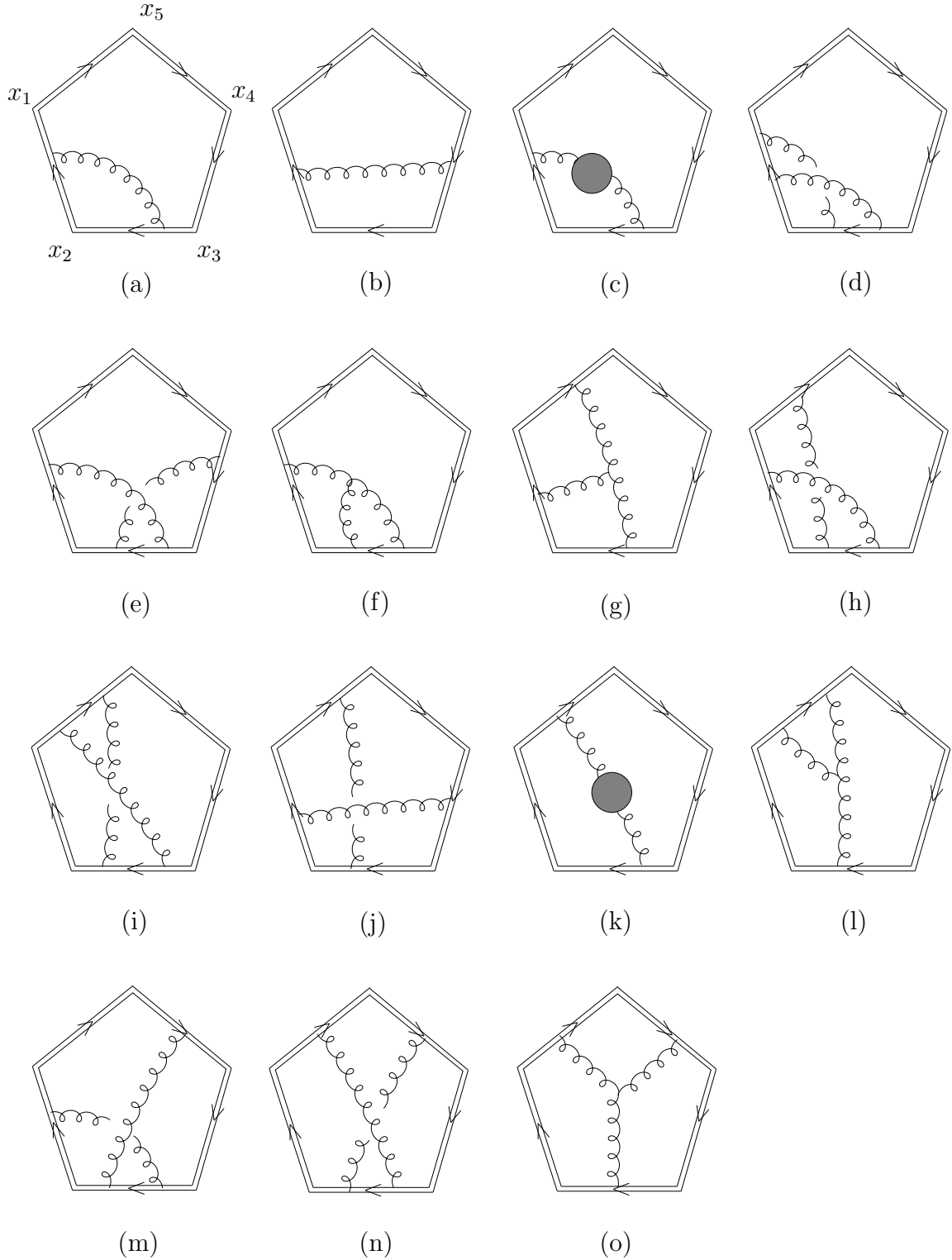}}
\caption[]{\small The Feynman diagrams contributing to $\ln {W(C_5)}$ to two
loops. The double line depicts the integration contour $C_5$, the wiggly line the
gluon propagator and the blob the one-loop polarization operator.}
\label{all-diags}
\end{figure}

To summarize, in Figs.~\ref{all-diags} (c) -- (o) we list all non-vanishing
two-loop diagrams of different topologies contributing to $w^{(2)}$. The diagrams
in Figs.~\ref{all-diags} (c) -- (l) have the same topology as for the rectangular
Wilson loop (see Ref.~\cite{Drummond:2007cf}), while the diagrams in
Figs.~\ref{all-diags} (m) -- (o) are specific to the pentagon Wilson loop
$W(C_5)$. The diagrams in Figs.~\ref{all-diags} (f), (g), (l) and (o) involve the
three-gluon interaction vertex of the $\mathcal{N}=4$ SYM Lagrangian. Their color
factors equal $C_F N$, and therefore they contribute to the function $w^{(2)}$ in
\re{W-decomposition}. The diagrams shown in Figs.~\ref{all-diags} (d), (e), (h),
(i), (j), (m) and (n) are non-planar and their color factors equal
$C_F(C_F-N/2)$. To identify their contribution to $w^{(2)}$, we have to retain
its maximally non-Abelian part only, that is, to replace their color factors by
$C_F(C_F-N/2)\to -C_F N/2$. Finally, the diagrams in Figs.~\ref{all-diags}(c) and
(k) involve the one-loop correction to the gluon propagator with the blob
denoting gauge fields/gauginos/scalars/ghosts propagating along the loop. Their
color factors equal $C_F N$ and they contribute directly to $w^{(2)}$. To
preserve supersymmetry, we evaluate these diagrams within the dimensional
reduction (DRED) scheme.
The two-loop correction $w^{(2)}$ is given by the sum over the individual
diagrams shown in Fig.~\ref{all-diags} plus crossing symmetric diagrams.
To compute these diagrams, we employ the technique developed in
Refs.~\cite{K88,KK92}.

The result of our calculation can be summarized as follows. It is convenient to
expand the contribution of each diagram in powers of $1/\epsilon$ and separate
the UV divergent and finite parts as
\begin{equation}\label{para}
w^{(2)} =   \sum_{\alpha} \bigg\{\frac12\left(\frac1{\epsilon^4}A_{-4}^{(\alpha)}
+ \frac1{\epsilon^3}A_{-3}^{(\alpha)} + \frac1{\epsilon^2}A_{-2}^{(\alpha)} +
\frac1{\epsilon}A_{-1}^{(\alpha)}\right)\sum_{i=1}^5
\lr{{-x_{i,i+2}^2}\,{\mu^2}}^{2\vep} + A_0^{(\alpha)}\bigg\} + O(\epsilon)\,,
\end{equation}where the sum goes over the two-loop Feynman diagrams of different topologies
shown in Fig.~\ref{all-diags}(c)--(o) and the factor $1/2$ has been inserted for
later convenience. Here $A^{(\alpha)}_{-n}$ (with $0\le n \le 4$) are
dimensionless functions of the ratio of distances $x_{k,k+2}^2$ (with
$k=1,\ldots,5$). Making use of \re{para}, we parameterize the contribution of
each individual diagram to $w^{(2)}$ by the set of coefficient functions
$A^{(\alpha)}_{-n}$. We would like to stress that the contribution of each
individual diagram to $w^{(2)}$, or equivalently the functions
$A^{(\alpha)}_{-n}$, are gauge dependent and it is only their sum on the
right-hand side of \re{para} that is gauge invariant.

In our analysis, we have heavily used the results of the two-loop calculation of
the rectangular light-like Wilson loop $W(C_4)$~\cite{KK92,Drummond:2007cf}. The
expression for $W(C_4)$ has the same general form as $W(C_5)$,
Eqs.~\re{W-decomposition} and \re{para}, with the corresponding coefficient
functions $A^{(\alpha)}_{-n}$ known explicitly. There are no reasons to expect
\textsl{{a priori}} that $W(C_4)$ and $W(C_5)$ should be related to each other.
Still, as we will see in a moment, there exist remarkable relations between the
contributions of various diagrams to the two Wilson loops. As an example, let us
consider the vertex-like diagrams shown in Fig.~\ref{all-diags}(c),
\ref{all-diags}(d) and \ref{all-diags}(f). The corresponding Feynman integrals
depend only on the single distance $x_{13}^2$, and therefore they give the same
contributions to $W(C_4)$ and $W(C_5)$. In a similar manner, the contribution of
the diagram shown in Fig.~\ref{all-diags}(e) depends only on two distances,
$x_{13}^2$ and $x_{24}^2$, and as a consequence it is the same for $W(C_4)$ and
$W(C_5)$. At the same time, examining the diagram shown in
Fig.~\ref{all-diags}(g) we found that it gives the same contribution to $W(C_4)$
and $W(C_5)$ for the leading, $O(1/\epsilon^2)$ term but this is not true for the
$O(1/\epsilon)$ term, as well as for the finite $O(\epsilon^0)$ term.

Summarizing our results for $W(C_5)$, we find that the coefficient functions
entering the right-hand side of \re{para} are given by

\begin{itemize}

\item UV divergent $O(1/\epsilon^4)$ terms in \re{para} only come from the two Feynman diagrams
shown in Figs.~\ref{all-diags}(d) and \ref{all-diags}(f). They coincide with
those for the rectangular Wilson loop $W(C_4)$
\begin{equation}\label{A4}
A_{-4}^{\rm (d)}=-\frac1{16}\,,\qqqquad A_{-4}^{\rm (f)}= \frac1{16}
\end{equation}
\item  UV divergent $O(1/\epsilon^3)$ terms in \re{para} only come from the two Feynman diagrams
shown in Figs.~\ref{all-diags}(c) and \ref{all-diags}(f). They coincide with
those for the rectangular Wilson loop $W(C_4)$
\begin{equation}\label{A3}
A_{-3}^{\rm (c)}=\frac1{8}\,,\qqqquad A_{-3}^{\rm (f)}=-\frac1{8}
\end{equation}
\item   UV divergent $O(1/\epsilon^2)$ terms only come from the Feynman diagrams
shown in Figs.~\ref{all-diags}(c)-- \ref{all-diags}(g). They coincide with those
for the rectangular Wilson loop $W(C_4)$
\begin{equation}\label{A2}
A_{{-2}}^{\rm (c)}=\frac14\,,\qquad A_{{-2}}^{\rm (d)}=-{\frac
{\pi^2}{96}}\,,\qquad A_{{-2}}^{\rm (e)}=-\frac{\pi^2}{24} \,,\qquad
A_{{-2}}^{\rm (f)}=-\frac14+{\frac {5}{96}}\,{\pi }^{2}\,,\qquad A_{{-2}}^{\rm
(g)}= \frac{{\pi }^{2}}{48}
\end{equation}\item  UV divergent $O(1/\epsilon^1)$ terms come from the Feynman diagrams shown in
Figs.~\ref{all-diags}(c)--\ref{all-diags}(h),\ref{all-diags}(k) --
\ref{all-diags}(m) and \ref{all-diags}(o).

\item  Finite $O(\epsilon^0)$ terms come from all Feynman diagrams shown in
Figs.~\ref{all-diags}(c)--\ref{all-diags}(o).

The expressions for $A_{{-1},0}^{\rm (c)}$, $A_{{-1},0}^{\rm (d)}$,
$A_{{-1},0}^{\rm (e)}$ and $A_{{-1},0}^{\rm (f)}$ are the same as for the
rectangular Wilson loop~\cite{Drummond:2007cf}, while the remaining coefficients
$A_{{-1}}-$ and $A_{{0}}-$ are given by complicated functions of the distances
$x_{i,i+2}^2$. Instead of calculating each of them separately, below we determine
their total sum.
\end{itemize}

We would like to stress that the relations \re{A4}, \re{A3} and \re{A2} are not
specific to the pentagon Wilson loop. The same relations hold true for an
arbitrary $n-$gon light-like Wilson loop $W(C_n)$ (with $n\ge 4$). They
ensure that, independently of the number of light-like segments, the
$O(1/\epsilon^4)$ and $O(1/\epsilon^3)$ terms cancel inside $w^{(2)}$ in the sum
of all diagrams leading to
\begin{equation}\label{w2-polygon}
w^{(2)} =\left\{\epsilon^{-2}\frac{\pi^2}{96}+\epsilon^{-1} \frac12\sum_\alpha
A_{-1}^{(\alpha)} \right\}\sum_{i=1}^n (-x_{i,i+2}^2\,\mu^{2})^{2\epsilon}+
O(\epsilon^0)\,.
\end{equation}Substituting this relation into \re{W-decomposition}, we find that it agrees with
the expected structure of UV divergences of a light-like Wilson loop in planar
$\mathcal{N}=4$ SYM theory,
\begin{align} \notag
\ln W(C_n) &=a \, w^{(1)} + 2 a^2\,  w^{(2)}  + O(a^3)
\\ \label{W_n-expected}
& = -\frac{1}{4}\sum_{l=1}^\infty a^l\left(\frac{\Gamma_{\rm
cusp}^{(l)}}{(l\epsilon)^2} + \frac{\Gamma^{(l)}}{l\epsilon} \right)\sum_{i=1}^n
(-x_{i,i+2}^2\,\mu^{2})^{l\epsilon}+ O(\epsilon^0)\,.
\end{align}
Here $a=g^2N/(8\pi^2)$ is the 't Hooft coupling, $\C_{\rm cusp}(a) =
\sum_{l=1}^{\infty} a^l \C^{(l)}_{\rm cusp}$ is the cusp anomalous dimension and
$\C(a) = \sum_{l=1}^{\infty} a^l \C^{(l)}$ is the collinear anomalous dimension
given by \re{cusp-2loop}. Furthermore, comparing \re{w2-polygon} with
\re{W_n-expected}, we deduce that the coefficient in front of $1/\epsilon$ on the
right-hand side of \re{w2-polygon} should be equal to
\begin{equation}\label{A-minus}
\sum_\alpha A_{-1}^{(\alpha)} =\frac{7}{8} \zeta_3\,.
\end{equation}This relation is extremely non-trivial, given the fact that each individual term
in the sum $A_{-1}^{(\alpha)}$ in general depends both on the number of segments
$n$ and on the distances $x_{jk}^2$ between the cusp points on the contour $C_n$.
For the rectangular Wilson loop, the relation \re{A-minus} has been verified in
Ref.~\cite{Drummond:2007cf}. In the next section we show that it also holds true
for arbitrary $n\ge 5$.

\subsubsection{Structure of the simple poles}

As was already mentioned, the coefficients $A_{-1}^{(\alpha)}$ of the simple
poles in the right-hand side of \re{para} are functions of the distances
$x_{jk}^2$. In the simplest case of the rectangular Wilson loop $W(C_4)$ they can
be expressed in terms of polylogarithm functions. Going over to the pentagon and,
in general, to $n-$gon light-like Wilson loops, we should expect to encounter
even more complicated expressions for the coefficients $A_{-1}^{(\alpha)}$.
However, having the relation \re{A-minus} in mind, we should also expect to find
a dramatic simplification in the sum over all diagrams. This suggests considering
the total sum $\sum_\alpha A_{-1}^{(\alpha)}$ rather than analyzing each individual
term. The next step would be to uncover the mechanism responsible for the above
mentioned simplification which could eventually lead to \re{A-minus}.
{In what follows, we describe the main steps of the calculation of the pentagon ($n=5$) Wilson loop and refer the interested reader to
\cite{Drummond:2008aq} for more details.}

To start with, let us introduce an auxiliary Feynman integral that will play a
crucial role in our analysis,
\begin{equation}\label{J-integral}
 {J(z_1,z_2,z_3) = - i  \left({\mu^2}{{\rm e}^{-\gamma}/\pi}\right)^{-\epsilon}\int
d^{4-2\epsilon} z\, G(z-z_1) G(z-z_2) G(z-z_3)}\,,
\end{equation}
with the gluon propagator $G(x)$ defined in \re{propagator}. Obviously,
$J(z_1,z_2,z_3)$ is a symmetric function of the three points $z_i^\mu$ (with
$i=1,2,3$) in Minkowski space-time. In what follows, we will need its value in
the special limit when two of the points are separated by a light-like interval.
Assuming that $z_{23}^2 \equiv (z_2-z_3)^2=0$, we find (see second reference in
\cite{K88})
\begin{equation}\label{J}
 {J(z_1,z_2,z_3) \stackrel{z_{23}^2=0}{=}  ( \mu^2{\rm e}^{-\gamma})^{2\varepsilon}
\frac{\Gamma(1-2\varepsilon)}{(2\pi)^4
4\varepsilon}\int_0^1 \frac{d\tau\, (\tau\bar
\tau)^{-\varepsilon}}{[-(\tau z_{21}+\bar \tau z_{31})^2]^{1-2\varepsilon}}}\,,
\end{equation}
where we introduced the notation for $\bar \tau = 1-\tau$ and $z_{jk}\equiv z_j
- z_k$. Notice that the integral \re{J} develops a single pole which has a simple
physical interpretation~\cite{K88}. It originates from the integration in
\re{J-integral} over $z^\mu$ approaching the light-like direction defined by the
vector $(z_2-z_3)^\mu$, so that the distances $(z-z_2)^2$ and $(z-z_3)^2$ vanish
simultaneously and the two propagators on the right-hand side of \re{J-integral}
become singular.

\begin{figure}[h]
\psfrag{x1}[cc][cc]{$x_3$} \psfrag{x2}[cc][cc]{$x_2$} \psfrag{x3}[cc][cc]{$x_1$}
\psfrag{x4}[cc][cc]{$x_5$} \psfrag{x5}[cc][cc]{$x_4$}
\psfrag{a}[cc][cc]{(a)} \psfrag{b}[cc][cc]{(b)} \psfrag{c}[cc][cc]{(c)}
\psfrag{d}[cc][cc]{(d)} \psfrag{e}[cc][cc]{(e)} \psfrag{f}[cc][cc]{(f)}
\psfrag{g}[cc][cc]{(g)} \psfrag{h}[cc][cc]{(h)} \psfrag{i}[cc][cc]{(i)}
\psfrag{j}[cc][cc]{(j)} \psfrag{k}[cc][cc]{(k)} \psfrag{l}[cc][cc]{(l)}
\psfrag{m}[cc][cc]{(m)} \psfrag{n}[cc][cc]{(n)} \psfrag{o}[cc][cc]{(o)}
\centerline{{\epsfysize5cm \epsfbox{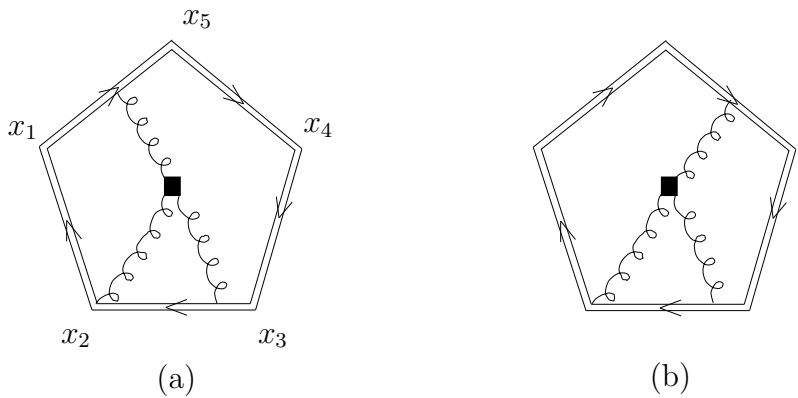}}} \caption[]{\small The
auxiliary Feynman diagrams defined in \re{I-aux}. The box depicts the fake
three-gluon vertex \re{J-integral}.} \label{Fig:aux}
\end{figure}

Let us now interpret the integral $J(z_1,z_2,z_3)$ as defining a new fake
`interaction vertex' for three gluons and examine the auxiliary Feynman diagrams
involving this vertex with all three points $z_i^\mu$ attached to the integration
contour ${C_5}$  as shown in Fig.~\ref{Fig:aux}. The reason for introducing these
diagrams is that, as we will see later in this section, they describe the simple
pole contribution to the `genuine' two-loop Feynman diagrams shown in
Fig.~\ref{all-diags}. Notice that one of the gluons in Fig.~\ref{Fig:aux} is
attached to the vertex of the pentagon, $z_2=x_2$, while the positions of the two
remaining gluons are integrated over the adjacent and the remaining non-adjacent
light-like segments, $z_3=x_3 + p_2 \tau_2$ and $z_1=x_{i+1}+p_i \tau_i$ (with
$i=4,5$). By definition, the Feynman diagrams associated with the two diagrams
shown in Fig.~\ref{Fig:aux} are
\begin{align}\label{I-aux}
 {I_{\rm aux}^{\rm (a)}} &= \frac12 g^4 C_F N (p_2 \cdot p_5) \int_0^1 d\tau_2 \int_0^1 d\tau_5\, J(x_1+p_5
\tau_5,x_2, x_3 + p_2 \tau_2)\,,
\\ \notag
 {I_{\rm aux}^{\rm (b)}} &=\frac12 g^4 C_F N (p_2 \cdot p_4) \int_0^1 d\tau_2 \int_0^1 d\tau_4\, J(x_5+p_4
\tau_4,x_2, x_3 + p_2 \tau_2)\,,
\end{align}
with $p_i = x_i - x_{i+1}$ being light-like vector, $p_i^2=0$. Substituting
\re{J} into \re{I-aux}, we obtain expressions for $I_{\rm aux}^{\rm (a,b)}$ in
the form of a three-fold integral. According to \re{J}, the integrals $I_{\rm
aux}^{\rm (a,b)}$ have a single pole in $1/\epsilon$. Then, we add together the
crossing symmetric diagrams of the same topology as in Fig.~\ref{Fig:aux} and
expand their contributions in $\epsilon$ similar to \re{para}
\begin{align}\label{M0}
I_{\rm aux}^{\rm (a,b)} + \text{(cross-symmetry)} =  {\lr{\frac{g^2}{4\pi^2}}^2 C_F N \left[ \frac1{2\epsilon}M_{-1}^{\rm (a,b)}
\sum_{i=1}^5 \lr{{-x_{i,i+2}^2}\,{\mu^2}}^{2\vep} + M_{0}^{\rm (a,b)} + O(\epsilon)\right]}\,.
\end{align}
where $M_{-1}^{\rm (a,b)}$ and $M_{0}^{\rm (a,b)}$ are complicated functions of
the distances $x_{jk}^2$ defined by \re{I-aux} and \re{J}. We will not need their
explicit form for our purposes.

The crucial observation is that the coefficient functions $A_{-1}^{(\alpha)}$
corresponding to the various diagrams shown in Fig.~\ref{all-diags} can be
expressed in terms of the two functions $M_{-1}^{\rm (a)}$ and $M_{-1}^{\rm
(b)}$. More precisely, our analysis leads to the following expressions for the
coefficient functions accompanying the simple poles in the pentagon Wilson loop
$W(C_5)$:
\begin{align}\nonumber
& A_{{-1}}^{\rm (d)}=-\frac1{24}\zeta_3\,, & & A_{{-1}}^{\rm (e)}=\frac12
\zeta_3\,, & & A_{{-1}}^{\rm (c)}+A_{{-1}}^{\rm (f)}={\frac {7}{24}} \zeta_3\,,
\\ \nonumber
 & A_{-1}^{\rm (g)} =
-M_{-1}^{\rm (a)}+\frac18 \zeta_3\,, & & A_{-1}^{\rm (h)} = 2 M_{-1}^{\rm (a)}\,,
  & & A_{-1}^{\rm (k)} +
A_{-1}^{\rm (l)} = -M_{-1}^{\rm (a)}-M_{-1}^{\rm (b)}\,,
\\ \label{A-1}
&  A_{-1}^{\rm (m)} = 2 M_{-1}^{\rm (b)}\,,   & &
 A_{-1}^{\rm (o)} = -M_{-1}^{\rm (b)}\,. & &
\end{align}
Putting these functions together we find that the functions $M_{-1}^{\rm (a)}$
and $M_{-1}^{\rm (b)}$ cancel in the sum of all diagrams! Most importantly, their
sum equals $\sum_\alpha A_{-1}^{(\alpha)}=\frac{7}{8} \zeta_3$, in perfect
agreement with \re{A-minus}.

The following comments are in order.

The relations \re{A-1} generalize similar ones for the rectangular Wilson loop
(see Eq.~(22) in Ref.~\cite{Drummond:2007cf}). The former relations can be
recovered by discarding the contribution of $M_{-1}^{\rm (b)}$ specific to the
pentagon contour and by identifying the function $M_{-1}^{\rm (a)}$ in the limit
$x_5=x_4$ with the corresponding function $\frac 18 M_2$ (see Eq.~(24) in
Ref.~\cite{Drummond:2007cf}).


It is straightforward to extend the analysis to arbitrary $n-$gon Wilson
loops $W(C_n)$. As was already explained, the two-loop corrections to $\ln
W(C_n)$, Eqs.~\re{w2-polygon} and \re{W_n-expected}, are described by the sum
$\sum_\alpha A_{-1}^{(\alpha)}$ receiving additional contributions from Feynman
diagrams of new topologies specific to the $n-$gon. Remarkably enough, the
coefficient functions $A_{-1}^{(\alpha)}$ can again be expressed in terms of the
auxiliary Feynman diagrams shown in Fig.~\ref{Fig:aux} in which the gluon coming
out of the box is attached to all possible light-like segments not adjacent to
the segments $[x_1,x_2]$ and $[x_2,x_3]$. Combining these diagrams together we
find that the single poles cancel in their sum leading to \re{A-minus} for any
$n$.

To summarize, we demonstrated by our explicit two-loop calculation that the
divergent part of an $n-$gon light-like Wilson loop has the expected form
\re{W_n-expected} with the cusp and collinear anomalous dimensions given by
\re{cusp-2loop}.

\subsubsection{Finite part}
We now turn to
the evaluation of the finite part of the pentagonal Wilson loop \re{para}. It receives contributions from the Feynman
diagrams shown in Figs.~\ref{all-diags}(c)--\ref{all-diags}(o). As in the
previous section, instead of analyzing each of them separately, we will consider
the total sum. Furthermore, based on the prediction of the conformal Ward
identities \re{remarkable} we expect that
\begin{equation}\label{A0-sum}
\sum_{\alpha={\rm c,\ldots, o}} A_0^{(\alpha)} = \, \frac{\pi^2}{48} \sum_{i=1}^5
\ln
 \Bigl(\frac{x_{i,i+2}^2}{x_{i,i+3}^2}\Bigr) \ln
 \Bigl(\frac{x_{i+1,i+3}^2}{x_{i+2,i+4}^2}\Bigr) - r \frac{\pi^4}{144}\,.
\end{equation}
Notice that the constant term in this relation is not determined by the Ward
identities, so $r$ is an arbitrary factor. The reason why we wrote it in the form
$\sim\pi^4$ is that the first term in the right-hand side of \re{A0-sum} has
transcendentality $4$.{}\footnote{ We adopt the standard convention
in the literature, where logarithms have transcendentality $1$ and $\zeta_{n}$
has transcendentality $n$. } Based on our analysis of the rectangular Wilson
loop~\cite{Drummond:2007cf}, we expect that the constant term should have the
same transcendentality which implies that $r$ should be a rational number.

We recall that the contribution of the diagrams shown in
Figs.~\ref{all-diags}(c)--\ref{all-diags}(f) to the finite part of $\ln W(C_5)$
can be obtained from the similar expressions for the rectangular Wilson
loop~\cite{Drummond:2007cf}
\begin{align}\label{A0-analytic}
& A_{{0}}^{\rm (d)}=-\frac54\cdot{\frac {7}{2880}}\,{\pi }^{4} \,,\quad
\\ \notag
& A_{{0}}^{\rm (c)}+A_{{0}}^{\rm (f)}=\frac54\cdot{\frac {119 }{2880}}\,{\pi
}^{4}\,,\quad
\\ \notag
& A_{{0}}^{\rm (e)}=\frac{\pi^2}{96}\sum_{i=1}^5
\ln^2\lr{\frac{x_{i+1,i-1}^2}{x_{i,i-2}^2}} -\frac54\cdot \frac{19}{720}\pi^4\,,
\end{align}
where the combinatorial factor $\frac54$ accounts for the different number of
diagrams of the same topology contributing to the pentagon and rectangular Wilson
loops. The finite contribution to $w^{(2)}$ from the remaining diagrams shown in
Figs.~\ref{all-diags}(g)--\ref{all-diags}(o) has to be calculated anew (with the
exception of the diagram shown in Fig.~\ref{all-diags}(j) which factorizes
into a product of two one-loop integrals corresponding to the diagram shown in
Fig.~\ref{all-diags}(b)). Let us separate them into two groups according to their
behavior as $\epsilon\to 0$:
\begin{itemize}
    \item The diagrams shown in Figs.~\ref{all-diags}(i), (j), (n) are finite;
    \item The diagrams shown in Figs.~\ref{all-diags}(h), (k)-(m),(o) and in Fig.~\ref{all-diags}(g) have, respectively, simple
    and double poles in $\epsilon$.
\end{itemize}
For the second group of diagrams we have to carefully separate the divergent and
finite parts following the definition \re{para}. This can easily be done by
employing the following `subtraction procedure'. As was shown in the previous
section, the simple poles in these diagrams are described by the functions
$M_{-1}^{\rm (a,b)}$ corresponding to the auxiliary diagrams shown in
Fig.~\ref{Fig:aux}. Therefore, in order to compensate the simple poles in the
above mentioned diagrams it is sufficient to subtract from them the auxiliary
diagrams with the appropriate weights defined in \re{A-1}. Since the auxiliary
diagrams also generate a finite part $M_{0}^{\rm (a,b)}$, Eq.~\re{M0}, the
subtractions will modify the finite parts of the individual diagrams, $A_{0}^{\rm
(\alpha)} \to \widehat A_{0}^{\rm (\alpha)}$ with
\begin{align}\nonumber
 & \widehat A_{0}^{\rm (g)} = A_{0}^{\rm (g)} + M_{0}^{\rm (a)}\,, & &
 \widehat A_{0}^{\rm (h)} = A_{0}^{\rm (h)}-2 M_{0}^{\rm (a)}\,,
  & &
\\\nonumber
& \widehat A_{0}^{\rm (m)}= A_{0}^{\rm (m)} - 2 M_{0}^{\rm (b)}\,,   & & \widehat
A_{0}^{\rm (o)}= A_{0}^{\rm (o)}+M_{0}^{\rm (b)}\,, & &
\\\label{A0-subtracted}
&  \widehat A_{0}^{\rm (k)+ (l)}=A_{0}^{\rm (k)} + A_{0}^{\rm (l)} +M_{0}^{\rm
(a)}+M_{0}^{\rm (b)}\,. &
\end{align}
Still, it is easy to see that the total sum of diagrams remains unchanged,
\begin{equation}\sum_{\alpha} A_{0}^{\rm (\alpha)} = \sum_{\alpha } \widehat A_{0}^{\rm
(\alpha)}\,.
\end{equation}The main advantage in dealing with the subtracted Feynman diagrams is that, by
construction, they are free from UV divergences%
\footnote{For the diagram shown in Fig.~\ref{all-diags}(g) we have to perform the
additional subtraction of the  double pole defined in \re{A2}.} and, therefore,
can be directly evaluated in $D=4$ dimensions. In this way, we found that,
remarkably enough,
\begin{equation}
\widehat A_{0}^{\rm (k)+ (l)}=0\,,
\end{equation}
and obtained the functions $\widehat A_{0}^{\rm (\alpha)}$ (with
$\alpha=\text{g,h,m,o}$) in the form of convergent multiple integrals. Their
explicit expressions are lengthy and to save space we do not present them here.
  We remark that the same subtraction procedure can be straightforwardly applied to define
the finite part of the $n$-cusp two-loop Wilson loop in terms of convergent multiple integrals.

Having an integral representation for the sum of the functions $\sum_{\alpha={\rm
g,h,m,o}} A_{0}^{\rm (\alpha)}$ and explicit expressions for the remaining
functions \re{A0-analytic}, we are now in a position to test the relation
\re{A0-sum} and to determine the factor $r$. Instead of trying to simplify the
sum of complicated multiple integrals, we performed thorough numerical tests.
Namely, both sides of \re{A0-sum} depend on the ratios of the distances between
the vertices of the pentagon $C_5$
\begin{equation}\label{distances}
X=\{x_{13}^2, x_{14}^2, x_{24}^2, x_{25}^2, x_{35}^2\}\,.
\end{equation}
To test \re{A0-sum} it is sufficient to evaluate both sides of \re{A0-sum} for
certain numerical values of $X$ and then to compare the resulting numerical
values. To be insensitive to the value of the constant term $r\pi^4$, we
evaluated the sum $\sum_{\alpha={\rm c,\ldots, o}} A_0^{(\alpha)}$ for several
sets of distances $X$, and examined their difference. The results of our
numerical tests are summarized in Table~\ref{tab:A0}.
\begin{table}[th]
\begin{center}
\begin{tabular}{|c||c|c|c|c|c| }
\hline $\{x_{13}^2, x_{14}^2, x_{24}^2, x_{25}^2, x_{35}^2\}$ &
$\widehat A_{0}^{\rm (g)}$ & $\widehat A_{0}^{\rm (h)}$ &  $\widehat
A_{0}^{\rm (m)}$ & $\widehat A_{0}^{\rm (o)}$ &  $ \sum_{\alpha={\rm
c,\ldots, o}} A_0^{(\alpha)}$
\\
\hline   $\{0.23,1.32,0.28,0.72,1.57\}$ & $+1.6415$ & $-1.6209$ &
$-5.4799$ & $+6.1485$ & $+0.5440$
\\
\hline   $\{ 2.23, 1.32,0.28,0.72,1.57\}$& $+5.8793$ & $-7.8712$ &$-4.4920$ &
$+6.2236$ & $-1.6328$
\\
\hline   $\{ 1.34, 0.04,0.98,3.21,1.43\}$ & $+5.4853$ & $-8.3748$ &
$-5.2143$ &$+5.8084$ & $-2.4030$
\\
\hline $ \{ 2.43, 1.30, 0.03,1.41,1.49 \}$  & $+4.1626$ & $-7.8060$
& $-5.1487$ & $+5.8841$ & $-1.7502$
\\
\hline  $\{ 5.32, 0.42, 1.23,7.76,2.53 \}$  & $+3.5254$ & $-3.9424$
& $-5.9461$ & $+5.1958$ & $-1.0333$
\\
\hline
\end{tabular}
\end{center}
\caption{The coefficient functions \re{A0-subtracted} and their
total sum \re{A0-sum} evaluated for different sets of the distances
\re{distances}.} \label{tab:A0}
\end{table}
We found that our results for the functions $A_0^{(\alpha)}$ are in perfect
agreement with the first term in the right-hand side of \re{A0-sum}. Then we
determined the numerical value of the factor $r$ parameterizing the constant term
in \re{A0-sum} leading to \begin{equation}r = 0.99996... \,.
\end{equation}Assuming maximal transcendentality of the constant term, we expect
$r$ to be rational. Our calculation suggests that $r=1$.

We conclude that our result for the two-loop pentagon Wilson loop is in agreement
with the prediction \re{remarkable} based on the conformal Ward identities.

\section{Conclusions}

In this paper we have provided further evidence for the weak-coupling duality
between MHV planar gluon amplitudes and light-like Wilson loops. Let us summarize
the main arguments in favor of this conjecture.

The light-like Wilson loops with cusps have an intrinsic conformal symmetry which
is broken, in a controlled way, by the cusp anomalies. This gives rise to
anomalous conformal Ward identities which unambiguously fix the form of the
finite part of the polygonal Wilson loop for $n=4$ and 5, and reduce the freedom
in the dependence on the kinematic variables to a function of conformal
invariants for $n\geq6$. The main open question is whether we can further
restrict or even fully determine this function by evoking some additional
properties. A promising scenario is to impose the consistency condition that the
Wilson loop with $n$ cusps reduce to the one with $n-1$ cusps when one of the
cusps is `flattened'. This requirement is analogous to the collinear
limit for MHV planar gluon amplitudes, as discussed in \cite{Bern:1994zx}. It
consists in taking two adjacent gluons to be nearly collinear, thus relating the
$n-$gluon to the $(n-1)-$gluon amplitude. The combination of conformal symmetry
with these additional consistency restrictions may turn out to be powerful enough
to completely fix the functional form of the polygonal Wilson loop to all orders.
Even so, it will still contain some dynamically determined parameters like the
cusp and collinear anomalous dimensions. We know that the former can be
determined to all loops from a Bethe Ansatz
\cite{Belitsky:2006en,BES,Basso:2007wd} and its strong coupling limit matches the
string theory prediction~\cite{GPK,Frolov:2002av,Kr02,Roiban:2007dq}. It is not
unlikely that the remaining parameters are also determined from integrability. If
all of this is true, the light-like Wilson loop could be a very interesting
example of a soluble theory.

Let us now turn to the MHV planar gluon amplitudes. What reasons do we have to
believe that they are equivalent to the perturbative light-like Wilson loops? By
now we have a lot of `experimental' evidence to this effect.

First of all, the finite part of the following amplitudes has been calculated
\begin{itemize}
\item  four-gluon amplitude up to three loops \cite{bds05};
\item  five-gluon amplitude up to two loops \cite{5point};
\item  $n$-gluon amplitude at one loop \cite{Bern:1994zx}.
\end{itemize}
In all these cases the results match those for the Wilson loops, found either by
explicit calculations (\cite{DKS07}, \cite{BHT07}, \cite{Drummond:2007cf} and the
present paper) or predicted by conformal invariance, as we have explained in
\cite{Drummond:2007cf} and in the present paper.

Secondly, the gluon amplitudes have the intriguing property of `dual' conformal
symmetry \cite{Drummond:2006rz} (see also \cite{DKS07,Nguyen:2007ya}). All the
scalar Feynman integrals appearing in the calculations of Bern et al up to four
loops\footnote{Taking dual conformal symmetry as an assumption, in
\cite{Bern:2007ct} a five-loop four-gluon amplitude was constructed, which meets
all the available unitarity consistency requirements. } are dual to conformal
integrals, after we take them off shell and perform the change of variables
\p{0002} from momenta to `dual coordinates'.\footnote{It should be pointed out
that exactly the same change of variables appears in the `T-duality'
\cite{Kallosh:1998ji,McGreevy:2007kt,Ricci:2007eq} transformation employed by
Alday and Maldacena \cite{am07} in their strong coupling treatment of the gluon
amplitudes.}
This makes us believe that the dual
conformal symmetry is not an accidental property of the gluon amplitudes, and
that its breaking is controlled by the same type of anomalies as for the Wilson
loop. In our opinion, investigating the origin of this symmetry is an important
task.

On the other hand, we know that dual conformal symmetry cannot uniquely fix the
gluon amplitudes with six or more gluons, just as in the case of a Wilson loop
with more than five cusps. As mentioned earlier, the additional input may come
from the collinear limit behavior studied in \cite{Bern:1994zx} and whose
relevance was reiterated in \cite{Bern:2006ew}. We believe that the chances for
this scenario to work out improve if one adds the requirement that the unknown
function in the case $n\geq6$ depend only on conformal invariants.

However, we should also be prepared to face a few less optimistic possibilities.
Even if dual conformal invariance is a true symmetry of the gluon amplitudes and
it restricts them in the way the intrinsic conformal invariance of the Wilson
loop does, it may be that the additional requirements (collinear limits for gluon
amplitudes and their equivalent for Wilson loops) are not enough to fully
determine the amplitude. This may happen at a relatively high loop order, where
the variety of Feynman integrals and of associated functions is rich enough and
one might be able to construct more than one function meeting all requirements.
In this case gluon amplitudes and Wilson loops  would fully agree for $n=4,5$,
but they would start deviating from each other for $n\geq6$ at two or higher
loops.

Another possibility is that dual conformal symmetry is an accidental property of
the low-loop gluon amplitudes. In this case, starting at some higher perturbative
order, even the four- or five-gluon amplitudes might loose their simplicity and
become similar to the very complicated generic QCD amplitudes.

All of these scenarios should be confronted with the discrepancy between the
large $n$ limit of the BDS conjecture and the strong coupling results recently
found by Alday and Maldacena \cite{Alday:2007he}. It may be that at strong
coupling one sees the entire perturbative expansion, with all possible breakdowns
of the above mechanisms.

What should be done at present to shed more light on these interesting issues? In
our opinion, two parallel calculations should be carried out in the near future:
the six-gluon two-loop MHV amplitude and the corresponding Wilson loop with six cusps.
When the results become available, we may have the answers to some of the above
questions.

\section*{Acknowledgements}

We would like to thank Z.~Bern, L.~Dixon, A.~Gorsky, J.~Maldacena, N.~Nekrasov,
J.~Plefka and V.~Smirnov for stimulating discussions. This research was supported
in part by the French Agence Nationale de la Recherche under grant
ANR-06-BLAN-0142.


\end{document}